\documentclass[a4paper,fleqn,usenatbib]{mnras}

\usepackage{graphicx}   
\usepackage{amsmath}    
\usepackage{amssymb}    

\usepackage[T1]{fontenc}
\usepackage{ae,aecompl}

\title[The rotation of Galactic GCs]{The eye of Gaia on globular clusters kinematics: internal rotation}
\author[Sollima et al.]{A. Sollima$^{1}$\thanks{E-mail:
antonio.sollima@inaf.it}, 
H. Baumgardt$^{2}$, M. Hilker$^{3}$\\
$^{1}$ INAF Osservatorio di Astrofisica e Scienza dello spazio di Bologna, via Gobetti 93/3, 40129 Bologna, 
Italy\\
$^{2}$ School of Mathematics and Physics, The University of Queensland, St. Lucia, QLD 4072, Australia\\
$^{3}$ European Southern Observatory,Karl-Schwarzschild-Str. 2, 85748 Garching, Germany
}

\date{Accepted 2019 February 15. Received 2019 February 11; in original form 2018 October 25}

\pubyear{2018}

\begin{document}
\label{firstpage}
\pagerange{\pageref{firstpage}--\pageref{lastpage}}
\maketitle


\begin{abstract}
We derived the three-dimensional velocities of individual stars in a sample of 
62 Galactic globular clusters using proper motions from the second data 
release of the Gaia mission together with the most comprehensive set of line-of-sight 
velocities with the aim of investigating the rotation pattern of these stellar 
systems. We detect the unambiguous signal of rotation in 15 clusters at amplitudes 
which are well above the level of random and systematic errors. For these clusters, we 
derived the position and inclination angle of the rotation axis with respect to the line of 
sight and the overall contribution of rotation to the total kinetic energy budget. 
The rotation strengths are weakly correlated with the half-mass radius, the relaxation 
time and anticorrelated with the destruction rate, while no significant alignment of the rotation axes with the orbital poles has been observed. 
This evidence points toward a primordial origin of the systemic rotation in 
these stellar systems.
\end{abstract}


\section{Introduction}
\label{intro_sec}

Among old ($>10~Gyr$) stellar systems, globular clusters (GCs) are those 
with the largest ratio between age and half-mass relaxation time. A typical 
GC star completed hundreds of orbits within the cluster potential and the 
chance of interaction with another star, which changes its orbit, is significant. The effect of 
a large number of interactions is to randomize the directions of individual 
orbits and to lead toward a velocity distribution tending to a Maxwellian-like 
distribution. 
For this reason, GCs are the prototype of pressure-supported stellar system, in 
which the gravitational potential energy of the cluster is balanced by the 
kinetic energy residing in random motions.
On the basis of the above considerations, the kinematics of GCs has been widely
investigated through the comparison with isotropic non-rotating models
\citep[like e.g.][models]{king66} in a large number of past studies
\citep[][and references therein]{mclaughlin05,baumgardt17,henaultbrunet19}.

Nevertheless, deviations from isotropy can be present in the velocity 
distribution of GC stars in the form of ordered motions (i.e. rotation) and/or 
preferential orientation of the velocity ellipsoid (i.e. anisotropy). 
Neglecting such effects can affect the determination of dynamical
parameters of GCs like e.g. their masses \citep{sollima15}.
In particular, the presence of rotation has deep relevance
for the equilibrium of these stellar systems contributing to their kinetic energy
budget and possibly leading to a flattening of their shape in the direction parallel to 
the rotation axis \citep{wilson75}.

Rotation in GCs can either be relic of the initial conditions of these 
objects at the epoch of their formation or originate from the interaction with 
the Galactic tidal field within which they move \citep{keenan75,mapelli17}. 
Although the formation mechanism of GCs is still not 
completely understood, a possible picture of their birth environment is 
provided by theoretical simulations of star forming complexes. In these environments
turbulence-supported molecular clouds possess several clumps which
merge over a time-scale of a few Myr \citep{mapelli17}. 
The large-scale torques occurring during the hierarchical assemby of these clumps imprint
a significant rotation to the embedded cluster which is 
enhanced during its early collapse because of angular momentum conservation.
During the subsequent long-term evolution, two-body relaxation tends to erase
such a rotation pattern. The dampening of the rotation signal is additionally caused by
the ever-continuing mass-loss experienced by the cluster which carry away
angular momentum \citep{tiongco17}. Part of the original rotation can however
survive till the present-day and be observable in the velocity distribution of
GC stars. 
A certain degree of rotation can also develop in the cluster outskirts because
of tidal effects. Indeed, at large distance from the cluster centre Coriolis
force is directed inward/outward according to the direction of the stellar 
motion with respect to the systemic cluster orbit, with stars on prograde orbits
more easily expelled by the cluster \citep{henon70,vesperini14}. On the long term 
this effect produces a low-amplitude retrograde rotation whose axis is aligned 
with the cluster orbital pole and with a period synchronized with the cluster 
orbital period \citep{tiongco16,tiongco18}. The measure of rotation in a statistically 
meaningful sample of GCs is therefore crucial to study the efficiency and frequency of the 
above processes.

The main effect of rotation is a shift in the mean tangential motion along a
preferential axis. Such a shift reflects in the velocity distribution along all
the three components in proportions depending on the position angle and the
inclination with respect to the line of sight of the rotation axis.
Thus, a thorough analysis of rotation requires an estimate of all 
three velocity components. The lack of accurate proper motions has represented a
major issue in the analysis of rotation of GCs till recent years.
For this reason, the large majority of the studies
conducted in the past on this topic were performed for GCs which are under favorable 
projection conditions (i.e. in case of edge-on rotation), where rotation
leaves a detectable sinusoidal modulation of the mean line-of-sight velocity 
as a function of the azimuthal position of the stars (see Sect. \ref{lit_sec}). 
The most recent and comprehensive studies based on large samples of line-of-sight
velocities analysed a few tens of GCs \citep{lane10,bellazzini12,fabricius14,kimmig15,lardo15,kamann18,ferraro18}.
In these studies, only the rotation velocity projected along the line of sight 
was determined and no information on the actual inclination of the rotation 
axis could be derived. In spite of this limitation, these studies revealed 
correlations between the rotation strength and various general 
\citep[horizontal branch morphology, absolute magnitude,
metallicity;][]{bellazzini12,lardo15}, structural \citep[e.g. ellipticity;][]{fabricius14} 
and dynamical \citep[half-mass relaxation time;][]{kamann18} parameters.
On the other hand, recent studies based on the analysis of proper motions 
measured have been conducted for the most 
massive and nearby clusters $\omega$ Centauri \citep{vanleeuwen00,vandeven06,libralato18} and 47 Tucanae 
\citep{anderson03,bellini17} and NGC 6681 \citep{massari13}.

A revolution in this field is provided by the astrometric mission Gaia which
measures parallaxes and proper motions for $\sim10^{9}$ stars in both hemispheres
with accuracies $<30\mu as$ (corresponding to $<1.5$ km/s at a distance of 10
kpc) sampling also thousands of stars in the outer regions of all Galactic GCs
\citep{gaia18a}.
In a recent paper, \citet{bianchini18} used the data from the Gaia 2nd data 
release to investigate rotation in a sample of 51 GCs and detected a 3$\sigma$ 
significant evidence in 11 of them. For a sub-sample of 8 GCs with available
line-of-sight velocities they also provide an estimate of the inclination angle of the 
rotation axis with respect to the
line of sight. This last work constitutes the most
extensive survey for rotation in Galactic GCs in terms of accuracy and
completeness to date. They confirm the correlation between 
the relevance of rotation over random motions and the half-mass relaxation time.
On the other hand, their work is based on proper motions only, thus suffering  
a similar detection bias of studies based on line-of-sight velocities only. In particular,
while their analysis has an excellent sensitivity in detecting rotation in
the plane of the sky, a rotation along the line of sight would not leave any
significant signal in the proper motions domain. Moreover, the presence of
covariances and small-scale systematics in the Gaia proper motions
\citep{arenou18} enhance the chance of false detections. Finally, the lack of
line-of-sight velocity information does not allow to derive the actual rotational 
velocity and inclination of the rotation axis for all the GCs of their sample.

In this paper we correlate the proper motions from the Gaia 2nd data release
with the most extensive survey of line-of-sight velocities collected by
\citet{baumgardt18}.
This allows to derive 3D velocities for more than 42,000 stars in 62 Galactic
GCs which are used to search for any significant rotation signal among these
stellar systems. We introduce the observational material and the sample selection
in Sect \ref{obs_sec}. The method adopted to detect rotation in our sample is
described in Sect. \ref{met_sec}. Sect \ref{mod_sec} is devoted to the 
modelling of the observed kinematics of GCs with a significant signal of 
rotation. We analyse correlations with various general and dynamical parameters
in Sect. \ref{corr_sec}. Finally, we discuss and summarize our results in Sect
\ref{concl_sec}. 

\section{Observational material}
\label{obs_sec}

\begin{figure*}
 \includegraphics[width=\textwidth]{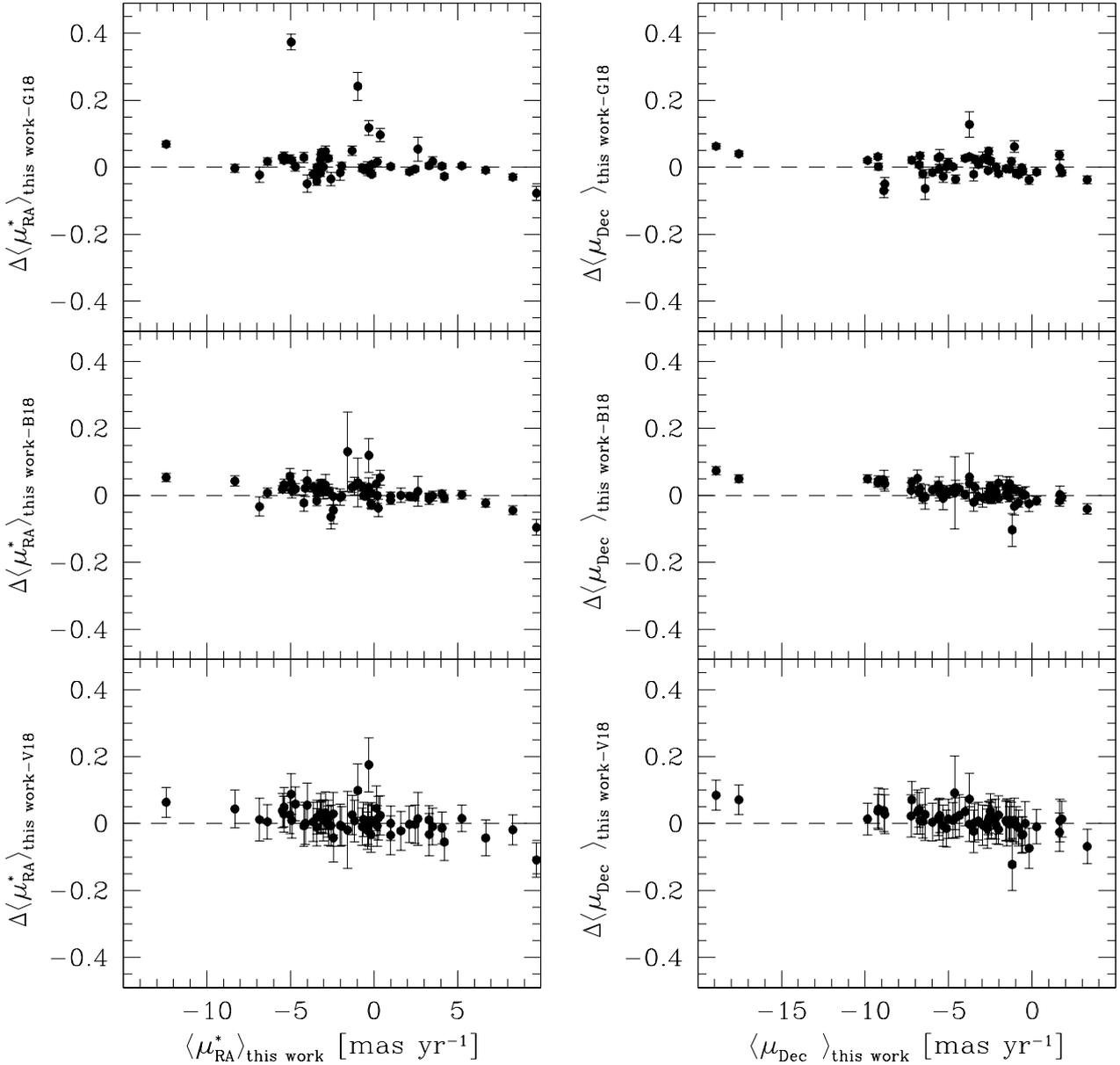}
 \caption{Comparison between the systemic proper motions estimated by 
 \citet{gaia18b}(top panels), \citet{baumgardt19}(middle panels), 
 \citet{vasiliev19}(bottom panels) and this work. The dashed
 line marks the one-to-one relation in all panels.}
\label{compsys}
\end{figure*}

The analysis performed in this paper is based on two main databases: {\it
i)} the sample of line-of-sight velocities collected by \citet{baumgardt18}, and {\it
ii)} the proper motions of the Gaia 2nd data release \citet{gaia18a}.
The line-of-sight velocity sample consists of 45,561 velocities measured in 109
Galactic GCs. It is a compilation of homogeneous measures from spectra obtained
at the Very Large Telescope of the European Southern Observatory and Keck 
telescope with different instruments, which
are complemented by published line-of-sight velocities in the literature. A
detailed description of the sample selection, reduction and data analysis
process as well as the complete list of references for the data resources 
is provided in \citet{baumgardt18} and \citet{baumgardt19}. 
The sample covers a wide portion of the
clusters' extent with a median uncertainty of 0.5 km/s.
  
For each target cluster we extracted from the Gaia public
archive\footnote{http://gea.esac.esa.int/archive/} the proper motions of all
stars within the tidal radius \citep[from][]{mclaughlin05}. We do not apply any
selection on either the Gaia quality flag, parallaxes or color-magnitude
diagram. Indeed, we found that any selection made on these basis provides only a
marginal improvement in terms of accuracy and purity against fore/background
contaminants while significantly reducing the sample size (see Sect. \ref{err_sec}). Proper motions
($\mu_{RA}~cos Dec$, $\mu_{Dec}$) and their corresponding uncertainties have
been converted into velocities ($v_{RA}$, $v_{Dec}$) using eq. 2 of \citet{gaia18b} assuming the distance listed in the \citet[2010
edition]{harris96} catalog and corrected for perspective rotation using eq.s 4
and 6 of \citet{vandeven06}. 

The two above samples have been cross-correlated providing 3D velocities for a
subsample of stars in each cluster. The celestial
coordinates (RA,Dec) have been converted into projected distances from the cluster centre
(X,Y) using eq.
1 of \citet{vandeven06} and adopting the centers of
\citet{goldsbury13}\footnote{for those clusters not included in the
\citet{goldsbury13} sample, we adopted the centers listed in the
\citet[][; 2010 edition]{harris96} catalog}.
The systemic motion of each cluster ($\langle v_{LOS}\rangle$, $\langle v_{RA}\rangle$, $\langle v_{Dec}\rangle$) has been determined by maximizing the
likelihood

\begin{eqnarray}
ln L&=&-\frac{1}{2}\sum_{i}\left[\frac{(v_{LOS,i}-\langle v_{LOS}\rangle)^{2}}{s_{LOS,i}^{2}}+
ln(s_{LOS,i}^{2} (1-\tilde{\rho_{i}}^{2}))+\right.\nonumber\\
& &\sum_{j=RA,Dec}\left(\frac{(v_{j,i}-\langle v_{j}\rangle)^{2}}{(1-\tilde{\rho_{i}}^{2}) s_{j,i}^{2}}+
ln(s_{j,i}^{2})\right)-\nonumber\\
& &\left.\frac{2 \tilde{\rho_{i}} (v_{RA,i}-\langle v_{RA}\rangle)(v_{Dec,i}-\langle
v_{Dec}\rangle)}{(1-\tilde{\rho_{i}}^{2}) s_{RA,i} s_{Dec,i}}\right]\nonumber\\
\label{lik_eq}
\end{eqnarray}

where

\begin{eqnarray*}
s_{j,i}^{2}&=&\sigma_{j,i}^{2}+\epsilon_{j,i}^{2}~~~~~~j=LOS,RA,Dec\nonumber\\
\tilde{\rho_{i}}&=&\rho_{i}\frac{\epsilon_{RA}\epsilon_{Dec}}{s_{RA} s_{Dec}}\nonumber\\
\end{eqnarray*}

In the above equation $\sigma_{j,i}$ and $\epsilon_{j,i}$ are the velocity dispersion 
and error of the i-th star in the j-th component (either LOS, RA or Dec) and 
$\rho_{i}$ is the covariance between $v_{RA}$ and $v_{Dec}$.
We neglected the covariaces involving RA and Dec (because of their negligible
amplitudes) and those involving parallax (since this parameter does not
enter in eq. \ref{lik_eq} and its uncertainty is therefore marginalized over
the other parameters).

The intrinsic velocity dispersion in any one direction at the projected distance of each star from the cluster
centre ($\sigma_{j,i}\equiv \sigma(R_{i})$; assumed the same in all the three
components) has been calculated by multiplying the 
amplitude of the line-of-sight velocity dispersion profile of the best fit 
\citet{king66} model provided by \citet{mclaughlin05} by the normalization 
factor providing the best fit to observational data.
We decided to normalize the amplitude of the velocity dispersion profile using
only line-of-sight velocities since proper motions have systematically larger
dispersions. This possibly arises from either an underestimate of the reported 
uncertainty or the possible presence of small-scale systematics (see also Sect.
\ref{err_sec}).
 
An iterative algorithm has been employed to calculate the mean velocities by
excluding at each iteration stars with
$$\sum_{j}\left(\frac{(v_{j,i}-\langle v_{j}\rangle)}{s_{j}}\right)^{2} > 25$$ 
and re-normalizing the $\sigma$ profile on the sample of retained stars.
The above algorithm generally converges after a few iterations providing the
systemic velocity along the three components, the amplitude of the intrinsic 
velocity dispersion profile and a sample of bona-fide cluster members.
In Fig. \ref{compsys} the estimated systemic proper motions are compared 
with the results by \citet{gaia18b,baumgardt19,vasiliev19}. We find an 
excellent agreement between our estimates and those provided by these works.

The sizes of the final samples range from 10 to
$\sim$2500 stars according to the cluster distance and mass.
Among the whole set of 109 GCs we selected only those 62 GCs with at least 50 
member stars with measurements in all the three velocity components.
We used these samples to search for rotation signals as described in Sect. 
\ref{met_sec}. 
For those clusters with positive
detections, in Sect. \ref{mod_sec} we fit dynamical models to extended samples containing all stars
with {\it at least} one measured velocity component and within 3$s_{j,i}$ from
the systemic velocity.
 
In the following sections we will refer to a system of coordinates in space 
(X,Y,Z) and velocity ($v_{X}$, $v_{Y}$, $v_{Z}$) where the coordinates of the
cluster centre and the systemic cluster motion have been subtracted

\begin{eqnarray*}
v_{X}&=&v_{RA}-\langle v_{RA}\rangle\nonumber\\
v_{Y}&=&v_{Dec}-\langle v_{Dec}\rangle\nonumber\\
v_{Z}&=&v_{LOS}-\langle v_{LOS}\rangle\nonumber\\
\end{eqnarray*}

\section{Method}
\label{met_sec}

\subsection{Algorithm}
\label{alg_sec}

\begin{table*}
 \centering
 \label{tab:table1} 
  \caption{Results for all the analysed GCs. 
  The classes are assigned to GCs in any of the following categories: r: rotating ($P>99.7\%$ and satisfying all the performed tests), 
  u: uncertain ($P>99.7\%$ but fails one of the tests against 
  random/systematic uncertainties; see Sect. \ref{err_sec}).}
  \begin{tabular}{@{}lrrr@{\hskip 1cm}lrrr@{\hskip 1cm}lrrr@{}}
  \hline
 Name    & $A$            & P     & class & Name     & $A$            & P     & class & Name     & $A$            & P     & class\\
         & $km/s$         & \%    &       &          & $km/s$         & \%    &       &          & $km/s$         & \%    &\\ 
 \hline
Arp 2    & 4.06 $\pm$4.15 &  67.6 &       & NGC 5904 & 4.11 $\pm$0.42 & 100.0 & r    & NGC 6522 & 1.43 $\pm$2.21 &  24.4 &    \\
NGC 104  & 5.00 $\pm$0.32 & 100.0 & r     & NGC 5927 & 0.93 $\pm$0.72 &  82.2 &      & NGC 6539 & 1.94 $\pm$1.15 & 95.9  &    \\
NGC 288  & 0.42 $\pm$0.32 &  84.3 & 	  & NGC 5986 & 1.65 $\pm$0.93 &  97.7 &      & NGC 6541 & 3.73 $\pm$1.15 & 100.0 & r  \\
NGC 362  & 0.51 $\pm$0.56 &  53.0 & 	  & NGC 6093 & 1.97 $\pm$0.84 &  99.7 & u    & NGC 6553 & 2.33 $\pm$0.82 & 100.0 & r  \\
NGC 1261 & 0.90 $\pm$0.64 &  86.9 & 	  & NGC 6121 & 0.22 $\pm$0.17 &  81.8 &      & NGC 6569 & 0.81 $\pm$0.91 &  52.3 &    \\
NGC 1851 & 0.45 $\pm$0.42 &  71.2 & 	  & NGC 6171 & 0.70 $\pm$0.46 &  93.9 &      & NGC 6624 & 0.87 $\pm$1.04 &  46.7 &    \\
NGC 1904 & 2.24 $\pm$0.46 & 100.0 & u	  & NGC 6205 & 1.53 $\pm$0.61 &  99.9 & r    & NGC 6626 & 2.42 $\pm$1.08 & 100.0 & r  \\
NGC 2808 & 2.25 $\pm$0.56 & 100.0 & r	  & NGC 6218 & 0.93 $\pm$0.37 & 100.0 & u    & NGC 6656 & 3.38 $\pm$0.71 & 100.0 & r  \\
NGC 3201 & 0.80 $\pm$0.41 &  98.8 & 	  & NGC 6254 & 0.26 $\pm$0.56 &  12.0 &      & NGC 6712 & 0.58 $\pm$0.69 &  45.8 &    \\
NGC 4372 & 1.36 $\pm$0.68 &  99.2 & 	  & NGC 6266 & 6.22 $\pm$1.53 & 100.0 & r    & NGC 6715 & 0.57 $\pm$1.11 &  18.5 &    \\
NGC 4590 & 0.27 $\pm$0.51 &  17.2 & 	  & NGC 6273 & 4.19 $\pm$1.12 & 100.0 & r    & NGC 6723 & 0.84 $\pm$0.55 &  93.0 &    \\
NGC 4833 & 1.14 $\pm$0.83 &  86.6 &       & NGC 6304 & 1.30 $\pm$0.91 &  89.4 &      & NGC 6752 & 0.91 $\pm$0.34 &  99.9 & u  \\
NGC 5024 & 1.17 $\pm$0.62 &  99.5 &  	  & NGC 6341 & 1.46 $\pm$0.61 &  99.9 & u    & NGC 6779 & 1.09 $\pm$1.52 &  35.1 &    \\
NGC 5053 & 0.36 $\pm$0.90 &  15.5 &  	  & NGC 6362 & 0.47 $\pm$0.40 &  75.2 &      & NGC 6809 & 0.88 $\pm$0.38 &  99.9 & u  \\
NGC 5139 & 4.27 $\pm$0.52 & 100.0 & r	  & NGC 6366 & 0.63 $\pm$0.60 &  66.6 &      & NGC 6838 & 0.74 $\pm$0.42 &  97.8 &   \\
NGC 5272 & 1.75 $\pm$0.42 & 100.0 & u	  & NGC 6388 & 1.51 $\pm$0.65 &  99.5 &      & NGC 7078 & 3.29 $\pm$0.51 & 100.0 & r \\
NGC 5286 & 0.76 $\pm$0.95 &  40.6 & 	  & NGC 6397 & 0.48 $\pm$0.17 & 100.0 & r    & NGC 7089 & 3.01 $\pm$0.70 & 100.0 & r \\
NGC 5466 & 0.84 $\pm$0.65 &  81.3 & 	  & NGC 6402 & 1.58 $\pm$0.88 &  97.7 &      & NGC 7099 & 1.10 $\pm$0.40 & 100.0 & u \\
NGC 5694 & 5.62 $\pm$6.40 &  62.3 & 	  & NGC 6440 & 3.93 $\pm$2.74 &  87.5 &      & Terzan 5 & 7.97 $\pm$2.38 & 100.0 & r \\
NGC 5824 & 6.47 $\pm$2.28 &  99.7 & u	  & NGC 6441 & 1.52 $\pm$1.66 &  55.3 &      & Terzan 8 & 0.79 $\pm$1.51 &  37.5 &   \\
NGC 5897 & 1.02 $\pm$0.79 &  82.0 & 	  & NGC 6496 & 1.21 $\pm$0.63 &  97.1 &      &          &                &       &   \\
\hline					    
\end{tabular}				    
\end{table*}				    
					    
\begin{figure}
 \includegraphics[width=8.6cm]{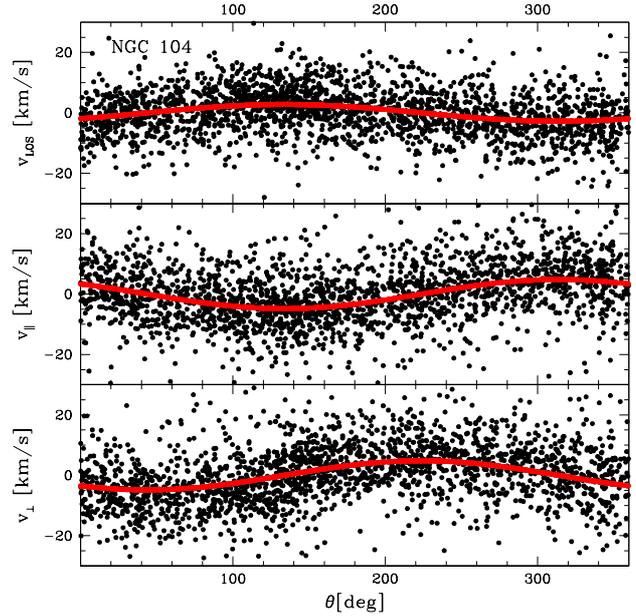}
 \caption{Distribution of the three velocity components as a function of the position angle for 
 NGC 104 (the entire set of best fits for the 15 GCs 
 with positive detection of rotation is available in the online version of the
 paper). The position angle is defined from North to West (see Sect. \ref{alg_sec}). The red solid lines show the best fit trend in all panels.}
\label{ang}
\end{figure}

The effect of rotation can be detected as a modulation of the mean velocity as a
function of the position angle. 
In particular, for a solid-body rotation with angular velocity $\omega$ 
the mean velocity in the three components is

\begin{eqnarray}
v_{Z}&=&\omega~R~sin(\theta-\theta_{0}) sin~i\nonumber\\
v_{\parallel}&=&\omega~R~sin(\theta-\theta_{0}) cos~i\nonumber\\
v_{\perp}&=&\omega~[R~cos(\theta-\theta_{0}) cos~i+Z sin~i]\nonumber\\
\label{transf_eq}
\end{eqnarray}

where $R$ is the projected distance from the cluster centre, $i$ is the
inclination angle of the rotation axis with respect to the line of sight,
$\theta$ is the position angle (defined anti-clockwise from the Y axis),
$\theta_{0}$ is the position angle of the rotation axis, $v_{\parallel}$ and
$v_{\perp}$ are the velocity components in the directions parallel and
perpendicular to the rotation axis, respectively (see Appendix A).

In a real cluster the angular velocity $\omega$ is a function of the distance
from the rotation axis. In particular, past studies found that rotating stellar 
systems show a solid-body rotation in their innermost region with an increasing
mean rotational velocity till a projected
distance where the rotation amplitude reaches a maximum ($\omega\sim
constant$ at $R<R_{peak}$) and then declines monotonically at larger distances
\citep{wilson75}. To account for this effect in a rigorous way a rotating model 
should be fitted to the data. However, this would introduce a dependence of the 
detection efficiency on the assumptions of the adopted model. 
To perform a model-independent analysis we considered an average projected rotation velocity 
$A=\langle\omega~R\rangle$ amplitude which has been assumed to be independent 
on distance.
Note that this approximation does not introduce any bias in the estimate of the 
position angle and inclination of the rotation axis since these quantities are
calculated from the ratio of amplitudes in $v_{Z}$, $v_{\parallel}$ and
$v_{\perp}$ and the term $\omega~R$ 
appears as a multiplicative factor in all the three eq. \ref{transf_eq}. So, 
as long as the same stars are used to compute the mean amplitudes in the three
components, any change in the $\omega~R$ factor of individual stars erases when 
relative amplitudes are computed.

In principle, the first two equations of \ref{transf_eq} allow to determine the 
values of $\theta_{0}$, $i$ and $A$. However, part of the information is lost in
neglecting the third equation which can increase the detection efficiency
while reducing the chance of false detections.
The third equation is complicated due to the presence of
the unknown distance along the line of sight (Z) as an independent variable. 
The dependence on Z does not affect the mean trend of $v_{\perp}$ (since $\langle
Z\rangle=0$) but introduces an additional spread in $v_{\perp}$ equal to
$\sigma_{\perp,Z}=\sigma_{Z}~A~sin i/R$, where $\sigma_{Z}$ is the spread in $Z$ 
as a function of the projected distance $R$. Although $\sigma_{\perp,Z}$ can be
calculated from the cluster density profile and the first two
equations of \ref{transf_eq}, we decided to neglect this spread. Indeed, the
term $\sigma_{Z}/R$ is negligible outside the innermost cluster region. 
On the other hand, at such a small distance the assumption of constant $A$ is 
no more reliable and in real clusters we expect $A\propto R$ implying small 
values of $\sigma_{\perp,Z}$ also in this region.

For each cluster of our sample we searched for the values of $\theta_{0}$, $i$ and
$A$ which maximize the likelihood 

\begin{eqnarray}
ln L&=&-\frac{1}{2}\sum_{i}\left[\frac{(v_{Z,i}-\overline{v_{Z,i}})^{2}}{s_{Z,i}^{2}}+
ln(s_{LOS,i}^{2} (1-\overline{\rho_{i}}^{2}))+\right.\nonumber\\
&
&\sum_{j=\parallel,\perp}\left(\frac{(v_{j,i}-\overline{v_{j}})^{2}}{(1-\overline{\rho_{i}}^{2}) s_{j,i}^{2}}+
ln(s_{j,i}^{2})\right)-\nonumber\\
& &\left.\frac{2 \overline{\rho_{i}}
(v_{\parallel,i}-\overline{v_{\parallel,i}})(v_{\perp,i}-\overline{
v_{\perp,i}})}{(1-\overline{\rho_{i}}^{2}) s_{\parallel,i} s_{\perp,i}}\right]\nonumber\\
\label{lik2_eq}
\end{eqnarray}

where

\begin{eqnarray*}
s_{j,i}^{2}&=&\sigma_{j,i}^{2}+\epsilon_{j,i}^{2}~~~~~~j=LOS,\parallel,\perp\nonumber\\
\epsilon_{\parallel,i}^{2}&=&\epsilon_{RA,i}^{2}sin^{2}\theta_{0}+\epsilon_{Dec,i}^{2}cos^{2}\theta_{0}-2
\rho_{i}\epsilon_{RA,i}\epsilon_{Dec,i}sin\theta_{0}cos\theta_{0}\nonumber\\
\epsilon_{\perp,i}^{2}&=&\epsilon_{RA,i}^{2}cos^{2}\theta_{0}+\epsilon_{Dec,i}^{2}sin^{2}\theta_{0}+2
\rho_{i}\epsilon_{RA,i}\epsilon_{Dec,i}sin\theta_{0}cos\theta_{0}\nonumber\\
\overline{\rho_{i}}&=&\frac{1}{s_{\parallel,i}s_{\perp,i}}
\left[\frac{(\epsilon_{Dec,i}^{2}-\epsilon_{RA,i}^{2})}{2}sin2\theta_{0}+
\rho_{i}\epsilon_{RA,i}\epsilon_{Dec,i}cos2\theta_{0}\right]\nonumber\\
\overline{v_{Z,i}}&=&A~sin(\theta_{i}-\theta_{0}) sin~i\nonumber\\
\overline{v_{\parallel,i}}&=&A~sin(\theta_{i}-\theta_{0}) cos~i\nonumber\\
\overline{v_{\perp,i}}&=&A~cos(\theta_{i}-\theta_{0}) cos~i\nonumber\\
\end{eqnarray*}

In the above notation $0^{\circ}<i<90^{\circ}$ is defined with respect to the
line of sight, $0^{\circ}<\theta_{0}<360^{\circ}$ grows anti-clockwise from 
North to West and $A$ is
positive for clockwise rotation in the plane of the sky. 

The above algorithm returns bestfit values of $\theta_{0}$, $i$ and
$A$ for all the 62 GCs of our sample. 
An example of best fit of the velocity distribution in the three components is
 shown in Fig. \ref{ang} for NGC 104.

To evaluate the significance of rotation
we employed a Monte Carlo technique: for each cluster, $10^{4}$ mock
observations of a non-rotating system have
been simulated by randomly extracting velocities in the three components from
Gaussian functions centered at $v_{Z},v_{RA},v_{Dec}=0$ and with dispersion
equal to the local velocity dispersion $\sigma_{i}$ at the position of the real
stars. The measurement errors have been then added as Gaussian shift with amplitude 
equal to the observational uncertainties of real stars (including
covariances between $v_{RA}$ and $v_{Dec}$) and the bestfit value of the
rotation amplitude has been calculated using the same technique adopted for real data. 
The fraction of simulations with best fit
amplitudes smaller than the one obtained on real data gives the probability that
the observed rotation signal is not produced by fluctuations.
An additional set of Monte Carlo simulations has been run for the clusters with
positive detection of rotation assuming the
position-dependent velocity shift in the three velocity components predicted by
the best fit models (see Sect. \ref{mod_sec}). The standard deviations of the
derived parameters ($i$, $\theta_{0}$, $A$, $\xi$) have been assumed as
the corresponding uncertainties. They do not include the uncertainty on the 
adopted cluster heliocentric distance, which affects both the rotation amplitude 
and the inclination angle as a systematic error in the following way

\begin{eqnarray*}
A&=&\sqrt{A_{LOS}^{2}+A_{\mu}^{2}d^{2}}\nonumber\\ 
i&=&tan^{-1}\left(\frac{A_{LOS}}{A_{\mu}d}\right)\nonumber\\
\end{eqnarray*}
where $A_{LOS}$ and $A_{\mu}$ are the amplitude measured in the line-of-sight velocity 
and proper motion spaces, respectively.

We list the bestfit amplitudes and the corresponding rotation probabilities for 
the entire set of 62 GCs of our sample in Table 1.  

\subsection{Effect of sample selection, random and systematic errors in Gaia proper motions}
\label{err_sec}

\begin{table}
 \label{tab:table2}
  \caption{List of rotating GCs}
  \begin{tabular}{@{}lrrrr@{}}
  \hline
 Name & $A$    & $\theta_{0}$ & $i$       & $\xi$ \\
      & $km/s$ & deg          & deg        &       \\ 
 \hline
 NGC 104  & -5.00 $\pm$0.32 & 224.3 $\pm$ 4.6 & 33.6 $\pm$ 1.8 & 0.102 $\pm$0.003\\
 NGC 2808 & -2.25 $\pm$0.56 &  36.1 $\pm$ 8.4 & 88.5 $\pm$10.3 & 0.020 $\pm$0.005\\
 NGC 5139 &  4.27 $\pm$0.52 & 170.2 $\pm$ 7.6 & 39.2 $\pm$ 4.4 & 0.045 $\pm$0.002\\
 NGC 5904 &  4.11 $\pm$0.42 & 221.6 $\pm$ 6.0 & 42.6 $\pm$ 3.2 & 0.137 $\pm$0.011\\
 NGC 6205 & -1.53 $\pm$0.61 & 165.5 $\pm$14.2 & 85.9 $\pm$11.6 & 0.131 $\pm$0.032\\
 NGC 6266 &  6.22 $\pm$1.53 & 104.2 $\pm$46.1 & 15.0 $\pm$12.8 & 0.043 $\pm$0.006\\
 NGC 6273 &  4.19 $\pm$1.12 &  56.9 $\pm$13.2 & 41.9 $\pm$ 7.1 & 0.065 $\pm$0.010\\
 NGC 6397 & -0.48 $\pm$0.17 &   8.6 $\pm$15.6 & 72.8 $\pm$11.9 & 0.004 $\pm$0.001\\ 
 NGC 6541 & -3.73 $\pm$1.15 &  83.2 $\pm$18.3 & 65.4 $\pm$13.9 & 0.083 $\pm$0.027\\
 NGC 6553 &  2.33 $\pm$0.82 & 237.7 $\pm$38.4 & 75.6 $\pm$29.5 & 0.003 $\pm$0.007\\
 NGC 6626 & -2.42 $\pm$1.08 &  28.6 $\pm$17.7 & 83.5 $\pm$13.3 & 0.018 $\pm$0.011\\
 NGC 6656 &  3.38 $\pm$0.71 & 252.8 $\pm$ 9.2 & 62.1 $\pm$ 6.3 & 0.091 $\pm$0.008\\
 NGC 7078 &  3.29 $\pm$0.51 &  52.6 $\pm$28.8 & 15.4 $\pm$ 5.4 & 0.071 $\pm$0.009\\
 NGC 7089 & -3.01 $\pm$0.70 & 346.6 $\pm$12.1 & 52.9 $\pm$11.2 & 0.071 $\pm$0.015\\
 Terzan 5 &  7.97 $\pm$2.38 & 260.4 $\pm$48.5 & 26.9 $\pm$34.6 & 0.026 $\pm$0.007\\
\hline
\end{tabular}
\end{table}

We find a rotation signal at $>3\sigma$ significance level in 24 GCs (see Table 1). 
However, as stated in Sect.
\ref{obs_sec}, we found that the intrinsic dispersion of proper motions is 
systematically larger than that of line-of-sight velocities. This effect can indicate
an underestimate of the uncertainties of proper motions, the presence of
small-scale systematics or by a significant contamination from Galactic 
interlopers. In the following we will check if these effects can
produce false detections in our sample.

In principle, random fluctuations increase the spread of the velocity
distribution without affecting the mean trend produced by rotation. However, an
underestimate of random errors affects the results of the Monte Carlo technique
spuriously increasing the significance level of the rotation signal. 
To quantify such an effect, we run the Monte Carlo simulations for the GCs
with positive detection multiplying the proper motion errors by the
fudge factor required to match the intrinsic dispersion of line-of-sight velocities. With
this approach we confirm a significant rotation signal in 23 out of 24 GCs, with
only NGC 5824 being excluded from the sample.

A more subtle effect can be produced by systematic errors. It is indeed known
that the inhomogeneous sampling of Gaia is responsible for the presence of
systematic shifts in proper motions. These errors have amplitudes
$\mu<0.07~mas/yr$ and follow a patchy structure with both small- and large-scale
variations \citep{lindegren18}. The presence of an inhomogeneous
distribution of proper motions within the cluster field of view can produce an
azimuthal variation mimicking a spurious rotation in the plane of the sky.
Actually, if we convert the rotation velocity component on the plane of the sky 
into proper motion we find that 19 GCs among those with positive detections have 
amplitudes $<0.1~mas/yr$, with the exception of NGC 104, NGC 5139, NGC 6266 and 
Ter 5. However, to produce a spurious rotation signal, the amplitude of proper 
motion systematics must be accompained by a negligible amplitude in line-of-sight 
velocity at the same position angle.
To test this effect, we calculate rotation amplitudes in our Monte Carlo 
simulations only from line-of-sight velocities and fixing the position  
and inclination angles of the rotation axis to those measured in real clusters. 
In this case, only 15 GCs 
maintain a significant rotation signal. They are listed in Table 2 together with
their best fit rotation amplitudes, position and inclination angles. 
For the 8 GCs excluded by this last 
criterion (namely, NGC 1904, NGC 5272, NGC 6093, NGC 6218, NGC 6341, NGC 6752, NGC 6809 
and NGC 7099; hereafter referred as "uncertain") we cannot exclude the presence of 
a genuine rotation mainly in the 
plane of the sky with amplitudes below the typical level of Gaia systematics.
Nevertheless, in the following Sections we will conservatively consider only the
15 GCs with an unambiguous evidence of rotation. 

Another possible source of missing/false detection could be in principle due 
to the contamination from Galactic field stars or astrometric artifacts. Velocity 
gradients in the fore/background Galactic component surrounding our analysed 
GCs are indeed present and can potentially produce a spurious rotation signal. 
However, the 3D velocity selection criteria described in Sect. \ref{alg_sec} are 
extremely effective in selecting a genuine sample of bright member stars: a 
comparison with the \citet{robin03} Galactic model predicts a 
contamination $<0.2\%$ in all the GCs of our sample with the exception of those 
GCs immersed in the bulge for which a few percent contamination is possible close 
to their tidal radii, where only a few stars are sampled. Moreover, over the relatively small extent of our GCs 
($<1~deg$ for the most extended GCs in the halo and $<0.3~deg$ for bulge GCs), 
the Galactic rotation pattern translates into a non negligible signal only 
along the Galactic bar. 

Astrometric artifacts can also increase the noise in the $v-\theta$ plane thus reducing the 
detection efficiency.
Consider however that the detection efficiency is only marginally dependent on 
errors. This is because the rotation signal depends on the amplitude of the mean velocity. So, by including 
stars with a large (or slightly underestimated) error would increase the spread 
around the mean trend without affecting the amplitude.

As stated in Sect. \ref{obs_sec} we have not applied any selection criteria based on 
neither Gaia quality flags nor on the distance from the cluster centre. To test the 
effect of more strict member selection on our results, we also performed the analysis adopting two different selection criteria: 
{\it i)} we exclude stars 
outside half of the nominal tidal radius \citep[from][]{mclaughlin05} and those with a proper motion error exceeding 1.5 times the central velocity 
dispersion, and {\it ii)} select only stars included in the member list provided by \citet{gaia18b}. 
In both cases, we confirm the significant rotation in 
the same 15 GCs. Regarding the selection criterion {\it (i)}, we also confirm 
the uncertain signal in 8 out of 9 GCs listed in Table 1, with the exception of
NGC5824 for which the number of member stars reduces below the treshold set at 50 objects.
As already noticed in \citet{baumgardt19}, the application of a proper motion 
error selection criterion improves the agreement between the dispersion masured along the 
line-of-sight and the transverse directions.
The estimated inclination angles and rotation strengths agree within the 
combined uncertainties with those measured in the unselected sample, 
although in the selected sample 
the uncertainties are larger because of the reduced sample sizes. 
The application of the selection criterion {\it (ii)} excludes from our sample a small fraction $<2\%$ in all the 
analysed GCs with no significant effect in the resulting detection as well as 
in the estimated inclination angles and rotation strengths.

On the basis of the above test we conclude that, for the purpose of the present 
work, quality and radial selection criteria do not affect the detection 
efficiency while reducing the accuracy of the estimated quantities. 
In the following sections we report the results obtained with the unselected 
sample.

\subsection{Comparison with literature results}
\label{lit_sec}

\begin{table*}
 \label{tab:table3}
  \caption{Detection of rotation in GCs from literature results: checkmarks, crosses 
  and tildes mark positive, negative and 2$\sigma$ detections, respectively.
  Checkmarks within parenthesis mark those GCs with uncertain detections found in this work.
  References: \citet[][L10]{lane10}, \citet[][B12]{bellazzini12}, \citet[][F14]{fabricius14}, 
  \citet[][L15]{lardo15}, \citet[][K15]{kimmig15}, \citet[][K18]{kamann18},
  \citet[][F18]{ferraro18}, \citet[][G18]{gaia18b}, 
  \citet[][B18]{bianchini18}, \citet[][K18]{vasiliev18}. Only GCs withat least a 2$\sigma$ detection have been listed.
 }
  \begin{tabular}{@{}lrrrrrrrrrrr@{}}
  \hline
 Name     & L10          & B12          & F14          & L15          & K15          & K18          & F18          & G18          & B18          & V18          & this work\\
 \hline
 NGC 104  & $\checkmark$ & $\checkmark$ &              &              & $\checkmark$ & $\checkmark$ &              & $\checkmark$ & $\checkmark$ & $\checkmark$ & $\checkmark$\\
 NGC 288  & X            & X            &              &              & X            &              & $\checkmark$ &              & X  	         & X            & X\\
 NGC 362  &              &              &              &              & X            & $\sim$       & $\checkmark$ &              & X  	         & X            & X\\
 NGC 1261 &              &              &              &              &              &              & $\sim$       &              &              &              & X\\
 NGC 1851 &              & $\checkmark$ &              & $\checkmark$ &              & $\checkmark$ &              &              & $\sim$	 & X            & X\\
 NGC 1904 &              & X            &              &              &              &              & $\checkmark$ &              & X  	         &              & ($\checkmark$)\\
 NGC 2808 &              & $\checkmark$ &              & $\checkmark$ & X            & $\checkmark$ &              &              & X  	         & X            & $\checkmark$\\
 NGC 3201 &              & $\checkmark$ &              &              &              & $\checkmark$ & $\checkmark$ &              & $\sim$	 & X            & $\sim$\\
 NGC 4372 &              &              &              & X            &              &              &              &              & $\checkmark$ & $\sim$       & $\sim$\\
 NGC 4590 & X            & $\checkmark$ &              &              & X            &              &              &              & X  	         & X            & X\\
 NGC 5024 & X            & X            & $\checkmark$ &              & X            &              &              &              &		 &              & $\sim$\\
 NGC 5139 &              & $\checkmark$ &              &              &              & $\checkmark$ &              & $\checkmark$ & $\checkmark$ & $\checkmark$ & $\checkmark$\\
 NGC 5272 &              &              & $\checkmark$ &              & X            &              & $\checkmark$ & $\checkmark$ & $\checkmark$ & $\sim$       & ($\checkmark$)\\
 NGC 5286 &              &              &              &              &              &              &              &              & $\sim$	 & X            & X\\
 NGC 5466 &              &              &              &              & $\checkmark$ &              &              &              &		 &              & X\\
 NGC 5824 &              &              &              &              &              &              &              &              &              &              & ($\checkmark$)\\
 NGC 5904 &              & $\checkmark$ & $\checkmark$ &              & $\checkmark$ & $\checkmark$ &              & $\checkmark$ & $\checkmark$ & $\checkmark$ & $\checkmark$\\
 NGC 5927 &              &              &              & $\checkmark$ &              &              & $\sim$       &              & X            & X            & X\\
 NGC 5986 &              &              &              &              &              &              &              &              & X            & X            & $\sim$\\
 NGC 6093 &              &              & $\checkmark$ &              &              & $\checkmark$ &              &              & $\sim$       & X            & ($\checkmark$)\\
 NGC 6121 & $\checkmark$ & $\checkmark$ &              &              & $\sim$       & $\sim$       &              &              & $\sim$       & X            & X\\
 NGC 6171 &              & $\sim$       &              &              &              &              & $\checkmark$ &              & X            & X            & X\\
 NGC 6205 &              &              & $\checkmark$ &              &              &              &              &              & $\sim$       & X            & $\checkmark$\\
 NGC 6218 & X            & X            & $\checkmark$ &              & X            &              &              &              & X            & X            & ($\checkmark$)\\
 NGC 6254 &              & X            & $\checkmark$ &              &              & $\checkmark$ & $\sim$       &              & $\sim$       & X            & X\\
 NGC 6266 &              &              &              &              &              & $\checkmark$ &              &              & $\sim$       & $\checkmark$ & $\checkmark$\\
 NGC 6273 &              &              &              &              &              &              &              &              & $\checkmark$ & $\checkmark$ & $\checkmark$\\
 NGC 6293 &              &              &              &              &              & $\checkmark$ &              &              &              & X            &\\
 NGC 6341 &              &              & $\checkmark$ &              & $\sim$       &              &              &              & X            & X            & ($\checkmark$)\\
 NGC 6388 &              & $\checkmark$ &              &              &              & $\checkmark$ &              &              & X            & X            & X\\
 NGC 6397 &              & X            &              &              &              &              &              &              & $\sim$       & X            & $\checkmark$\\
 NGC 6402 &              &              &              &              & X            &              &              &              & $\sim$       & X            & $\sim$\\
 NGC 6441 &              & $\checkmark$ &              &              & X            & $\sim$       &              &              &              & X            & X\\
 NGC 6496 &              &              &              &              &              &              & $\sim$       &              &              &              & X\\
 NGC 6522 &              &              &              &              &              & $\sim$       &              &              & X            &              & X\\
 NGC 6539 &              &              &              &              &              &              &              &              & $\sim$       & X            & $\sim$\\
 NGC 6541 &              &              &              &              &              & $\checkmark$ &              &              & $\sim$       & X            & $\checkmark$\\
 NGC 6553 &              &              &              &              &              &              &              &              &              & X            & $\checkmark$\\
 NGC 6626 &              &              & $\checkmark$ &              &              &              &              &              & X            & X            & $\checkmark$\\
 NGC 6656 & $\checkmark$ & $\checkmark$ &              &              & X            & $\checkmark$ &              & $\checkmark$ & $\checkmark$ & $\checkmark$ & $\checkmark$\\
 NGC 6681 &              &              &              &              &              & $\sim$       &              &              & X            & X            &\\
 NGC 6715 &              & $\checkmark$ &              &              & X            &              &              &              &              & X            & X\\
 NGC 6723 &              &              &              &              &              &              & $\sim$       &              &              & X            & X\\
 NGC 6752 & X            & X            &              & X            & X            & $\sim$       &              & $\checkmark$ & $\checkmark$ & $\sim$       & ($\checkmark$)\\
 NGC 6779 &              &              & $\checkmark$ &              &              &              &              &              & $\sim$       & X            & X\\
 NGC 6809 & $\sim$       & $\sim$       &              &              & $\sim$       &              &              & $\checkmark$ & $\checkmark$ & $\sim$       & ($\checkmark$)\\
 NGC 6838 &              & $\sim$       &              &              & X            &              &              &              & X            & X            & $\sim$\\
 NGC 6934 &              &              & $\checkmark$ &              & X            &              &              &              &              &              &\\
 NGC 7078 &              & $\checkmark$ &              & $\checkmark$ & $\checkmark$ & $\checkmark$ &              & $\checkmark$ & $\checkmark$ & $\checkmark$ & $\checkmark$\\
 NGC 7089 &              &              &              &              & $\checkmark$ & $\checkmark$ &              &              & $\checkmark$ & $\checkmark$ & $\checkmark$\\
 NGC 7099 & X            & X            &              &              & X            & $\sim$       &              &              & X            & X            & ($\checkmark$)\\
 Terzan 5 &              &              &              &              &              &              &              &              &              &              & $\checkmark$\\
\hline
\end{tabular}
\end{table*}

\begin{figure*}
 \includegraphics[width=\textwidth]{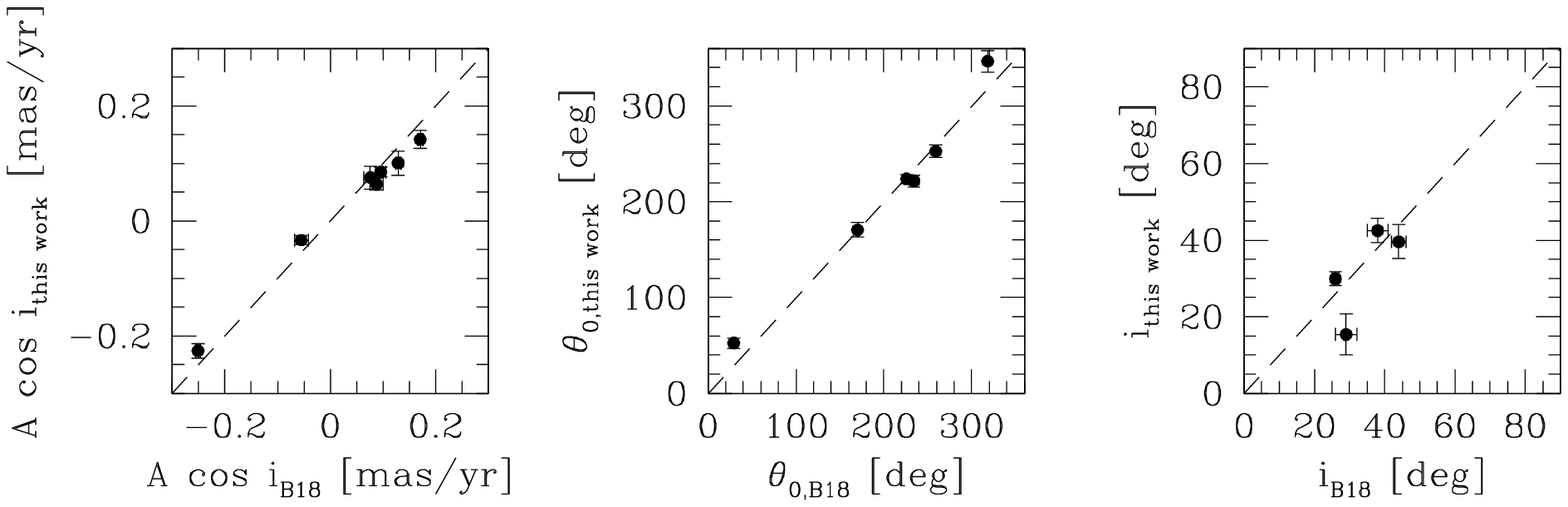}
 \caption{Comparison between the rotation amplitudes in the plane of the sky
 (left panel), position angle (middle panel) and inclination angle (right panel) 
 of the rotation axis estimated by \citet{bianchini18} and this work. The dashed
 line marks the one-to-one relation in all panels.}
\label{comp}
\end{figure*}

The results presented in this paper can be compared with those obtained by
previous groups in the past. In particular, a natural comparison can be made with
the work by
\citet{bianchini18} who detected rotations in 11 GCs using only proper 
motions form the same database adopted here. We restrict ourselves to those GCs
with a $3\sigma$ detection in both works. Seven GCs have been found to rotate in
both analyses (NGC 104, NGC 5139, NGC5904, NGC 6273, NGC 6656, NGC 7078 and NGC
7089).
Three GCs (NGC 5272, NGC 6752 and NGC 6809) also classified as rotating in both
analyses have been excluded in our sample because of their low-rotation 
amplitude in proper motion which could be mimicked by small-scale systematics. 
In this regard, \citet{bianchini18} claimed that systematics should be negligible
within the field of view covered by these clusters. However, as already reported
in Sect. \ref{err_sec}, although such a low-amplitude rotation cannot be
excluded for these GCs, we conservatively exclude them from our final sample.
We find only a 2$\sigma~(95.4\%<P<99.7\%)$ significant rotation signal in NGC 4372. This cluster
is the one with the smallest rotation amplitude in the sample of 
\citet{bianchini18}. Being at the border of the significance limits in both works,
the actual rotation in this cluster is not clear. 
NGC 6553 and Ter 5 are not included in the sample analysed of \citet{bianchini18}.
For three clusters (NGC 2808, NGC 6205 and NGC 6626) we measured inclination
angles larger than $i>80^{\circ}$ implying a rotation signal almost entirely
contained in the line-of-sight velocity space. Since the work by \citet{bianchini18} uses
only proper motions, they are insensitive to the rotation in the plane
perpendicular to the line of sight and it is therefore not surprising that they do not
detect any significant rotation in these GCs, although they report a 2$\sigma$ significance level for NGC 6205.
For the remaining three GCs (NGC 6266, NGC 6397 and NGC 6541) \citet{bianchini18}
found a 2$\sigma$ rotation signal. Note that our work, being based on all the 
three velocity components, has a higher detection efficiency than that of 
\citet{bianchini18} and thus is able to detect rotation also in these clusters.
The comparison between the estimated amplitudes, position and inclination axis
for the clusters in common is shown in Fig. \ref{comp}. A good agreement is
found for all these quantities.

The same conclusions hold for the rotation claimed by \citet{gaia18b} in the
Gaia Science verification paper, who find rotation in 8 GCs: 5 included in our final 
sample (NGC 104, NGC 5139, NGC 5904, NGC 6656 and NGC 7078) and three excluded 
in our analysis because of their low rotation 
amplitudes (NGC 5272, NGC 6752 and NGC 6809; see above).

Recently, \citet{vasiliev18} analysed Gaia proper motions in $\sim$80 Galactic GCs 
and studied their rotation pattern adopting stringent tresholds to account for 
the effect of systematic errors. We confirm the rotation detected by this author 
in the 8 GCs for which he states a 3$\sigma$ level detection. Of the 10 GCs with a 
2$\sigma$ level detection, five are 
classified as "uncertain" in our work (NGC 5272, NGC6093, NGC 6341, NGC 6752, 
NGC 6809), three show a 2$\sigma$ significant rotation (NGC 4372, NGC 5986, NGC 6341), one is 
found to be non-rotating (NGC 6388) and one is not included in our sample (IC 1276). 
As for the \citet{bianchini18} work, no rotation could be detected by \citet{vasiliev18} 
in the 6 rotating GCs with inclination angles $i>65^{\circ}$ (NGC 2808, NGC 6205, 
NGC 6397, NGC 6541, NGC 6553, NGC 6626), while Ter 5 is not included 
in his sample.

Among works based on line-of-sight velocities, 
\citet{lane10} analysed the kinematics of 10 GCs and found a significant
rotation in 3 of them. We find the same rotation signal in two of them (NGC
104 and NGC 6656) while we do not obtain a significant detection in NGC 6121.

\citet{bellazzini12} merged literature data with their own results for a sample of
24 Galactic GCs. Although they do not report a list of significant
rotators, 13 GCs of their sample have amplitudes at $3\sigma$ above the
statistical errors. We find significant rotation for 6 of them (NGC 104, NGC
2808, NGC 5139, NGC5904, NGC 6656, NGC7078), while we do not
confirm rotation in the remaining 7 (NGC 1851, NGC3201, NGC 4590, NGC 6121,
NGC 6388, NGC 6441 and NGC 6715), although NGC 3201 has a rotation signal at a 
2$\sigma$ significance level.
They do not find rotation in NGC 6397. However, for this cluster we find a very
small rotation amplitude in the radial component ($A~sin i\sim0.46~km/s$; i.e
similar to their reported measurement error). It is therefore not surprising
they were not able to detect such a small signal.

\citet{fabricius14} claimed the presence of rotation in all the 11 GCs of their 
sample. We confirm rotation for only 3 of them \citep[NGC 5904, NGC 6205 and NGC
6626 i.e. those with the largest rotation amplitude in][]{fabricius14}, while 4 
GCs (NGC 5272, NGC 6093, NGC 6218 and NGC 6341) were classified as "uncertain" 
in our analysis. 
It is worth stressing that rotation in the work by 
\citet{fabricius14} is measured as a mean slope in the velocity field derived
through integrated spectra of the cluster core. This technique can in principle
be affected by spurious detections due to the inhomogenous distribution of 
individual bright stars. While they report uncertainties
as small as $0.1~km/s~arcmin^{-1}$, it is possible that their 100\% detection
rate might be overestimated.

\citet{lardo15} analyse 7 GCs and detect significant rotation in 4
of them. We confirm their result for NGC 2808 and NGC 7078 (those with the
largest rotation amplitude in their sample), while we do not find any significant
signal in NGC 1851 and NGC 5927.

Similarly, \citet{kimmig15} detected rotation in 4 GCs in common with
our sample (NGC 104, NGC 5904, NGC 7078 and NGC 7089) and in NGC 5466 (for which
we do not confirm rotation). On the other hand, the rotation signal they 
measured in NGC 2808 and NGC 6656 was not significant in their analysis, likely
because of their small sample.

We confirm rotation in 9 among the 12 GCs in common with the study of 
\citet{kamann18} (NGC 104,
NGC 2808, NGC 5139, NGC 5904, NGC 6266, NGC 6541, NGC 6656, NGC 7078 and NGC
7089) while we classified as "uncertain" the detection in NGC 6093 and NGC 7099. 
We instead do not find 
any significant signal in NGC 1851. Note that the study by \citet{kamann18} uses several
thousands stars per cluster being
more sensitive than our study in those clusters with a small rotation amplitude
and aligned with the line of sight.
Moreover, the samples used by \citet{fabricius14} and \citet{kamann18} cover 
the innermost 30\arcsec of their GCs, while our work sample the clusters mainly outside 
their cores.

Finally, among the 6 GCs classified as rotating GCs by \citet{ferraro18} we 
find only a 2$\sigma$ significant rotation in NGC 3201 and classified 
as "uncertain" 2 of them (NGC 1904 and NGC 5272). No rotation has been instead found 
in NGC 288, NGC 362 and NGC 6171 in our analysis. Note that \citet{ferraro18} defines a 
detection when a significant asymmetry in the velocity distribution about the rotation 
axis is apparent in any radial bin, while they report no significant 
signal when the entire sample is considered (like in our analysis).

The comparison with the above literature results is summarized in Table 3.

Other studies devoted to individual clusters detected rotation in many GCs of our
final list: NGC 104
\citep{meylan86,gebhardt95,strugatskaya88,gerssen02,anderson03}, NGC
5139 \citep{woolley64,merritt97,norris97,vanleeuwen00,reijns06,vandeven06,pancino07,sollima09}, NGC 5904
\citep{lanzoni18}, NGC 6397
\citep{gebhardt95}, NGC 6656 \citep{peterson94}, NGC 7078
\citep{gebhardt94,dubath94,drukier98} and NGC 7089 \citep{pryor86}.
On the other hand, we detect a rotation at a 2$\sigma$ level in 
NGC 3201 \citep{cote95} and NGC 4372 \citep{kacharov14}, and an "uncertain" 
classification has been given to NGC 5272 \citep{kadla85}. 
We do not confirm instead the rotation found in NGC 5024 \citep{boberg17}, 
NGC 6388 \citep{lanzoni13} and NGC 6121 \citep{malavolta15}. 

\section{Dynamical modelling}
\label{mod_sec}

\begin{figure*}
 \includegraphics[width=\textwidth]{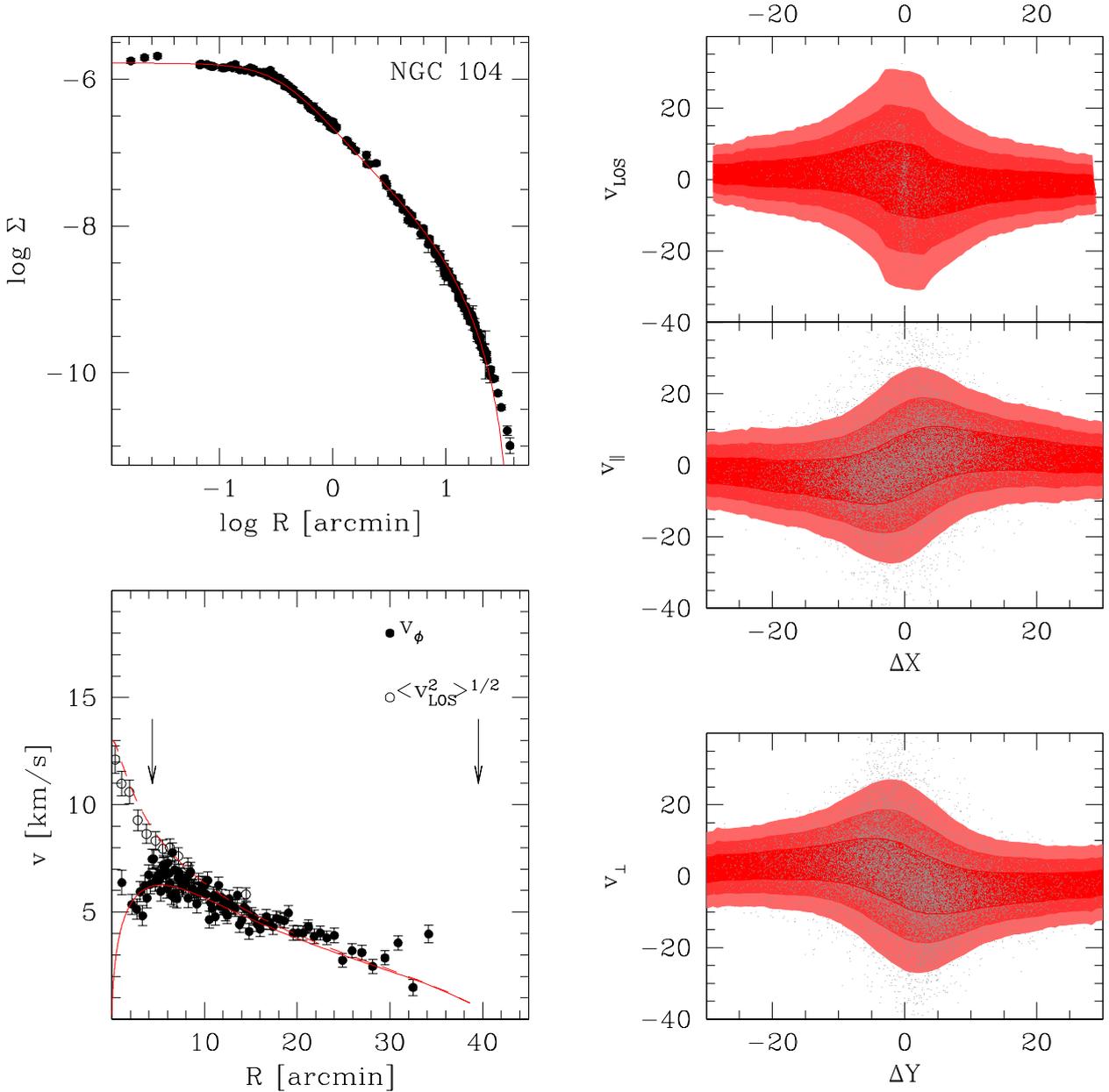}
 \caption{Best fit model of NGC 104 (the entire set of models for the 15 GCs 
 with positive detection of rotation is available in the online version of the
 paper). Top-left panel: projected density profile, Black points represent the 
 profile of \citet{trager95}. 
 Bottom-left panel: rotation 
 (solid line) and velocity dispersion (dashed line) profiles. Filled and open 
 dots represent the corresponding observed profiles. The location of the half-mass and tidal 
 radii are marked by arrows. 
 Right panels: distributions of velocities in 
 the three components as a function of the distance along ($\Delta Y$) and from 
 ($\Delta X$) the rotation axis. Red shaded area indicate the 1, 2 and 3$\sigma$ 
 intervals. Grey points mark observational data. For clarity, only velocities with 
 errors smaller than $5~km/s$ are plotted.}
\label{fit}
\end{figure*}

\begin{figure*}
 \includegraphics[width=\textwidth]{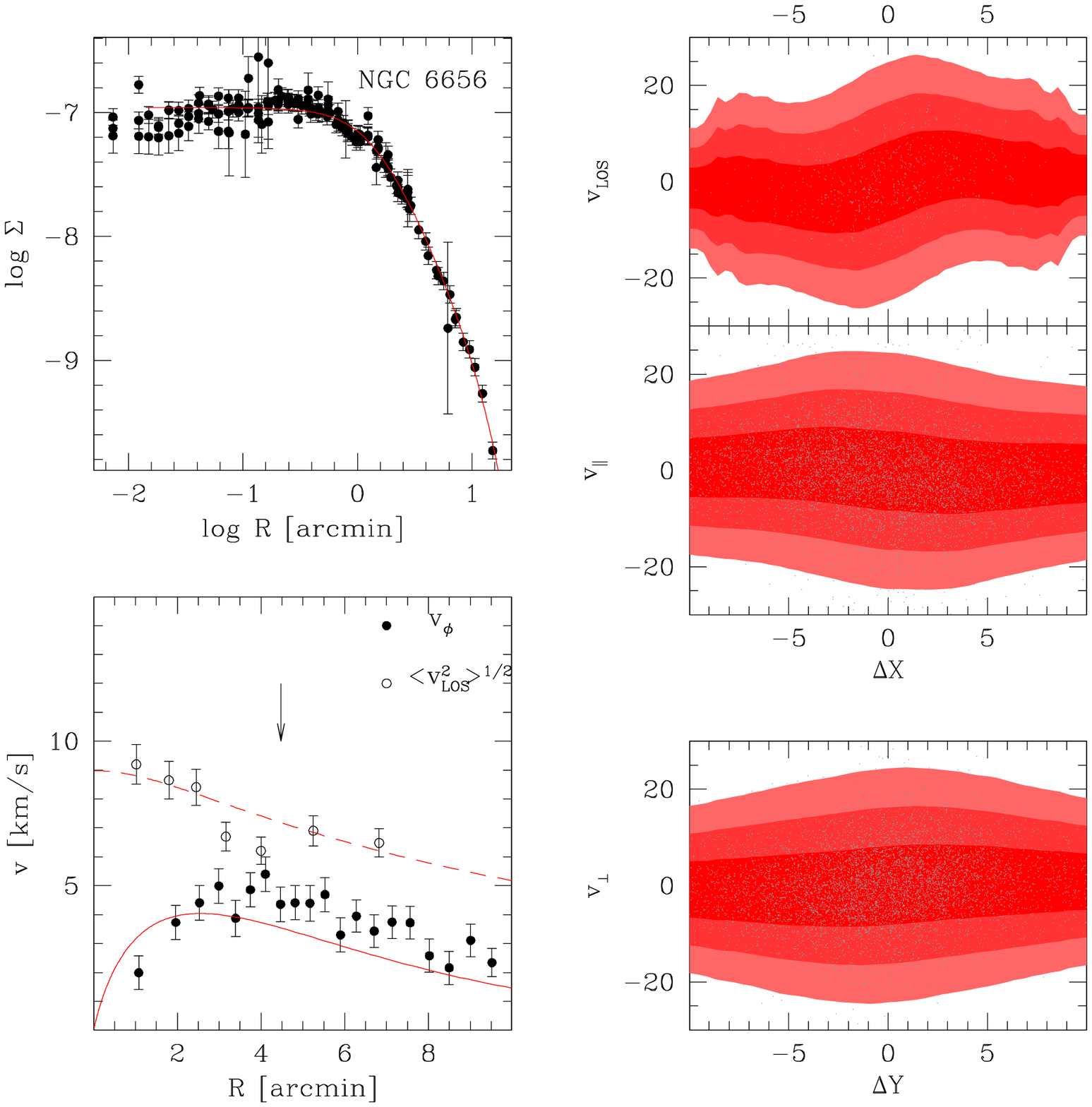}
 \caption{Same as Fig. \ref{fit} for NGC 6656.}
\label{fit2}
\end{figure*}

The algorithm described in Sect. \ref{alg_sec} provided a measure of the mean 
rotation amplitude in our sample of GCs. However, in a real cluster the rotation velocity depends on 
position. On the other hand, the efficiency of the radial 
sampling of our samples is given by the combination of the selection function 
of the considered databases. These depend on many factors like the heliocentric 
distance, the crowding 
conditions, etc. resulting in an inhomogeneous sampling of the clusters. 
Thus, the derived mean 
amplitudes are biased estimates of the actual rotation of our GCs and should 
not be compared with each other. A more appropriate way to derive the characteristics 
of the rotation of our GCs is to compare our dataset with suitable dynamical 
models. As specified in Sect. \ref{obs_sec}, for this comparison we used for each 
cluster the extended samples of stars i.e. those stars having at least one measure 
of their line-of-sight velocities or proper motions. Additionally, the surface brightness 
profiles of \citet{trager95} have been employed as further constrains.

Many self-consistent models of rotating stellar systems have been 
developed in the past \citep[e.g. ][]{wilson75}. These models are defined from a 
distribution function depending on two integrals of motions (energy and angular 
momentum along the rotation axis) which is integrated in a cylindrical 
coordinate system. Indeed, rotating stellar systems generally present a flattening 
along the direction of the rotation axis due to the kinetic energy excess along 
the radial direction in the equatorial plane produced by ordered motions. As a side effect,
both the projected density profile and the resulting velocity distributions depend 
on the same model parameters. 
Unfortunately, the adopted set of structural and 
kinematic data have very different accuracies with the fractional 
uncertainty in the surface brightness being about one order of magnitude 
smaller than that in the velocity dispersion in the same radial intervals.
As a consequence, the fit of the dynamical model is entirely dominated by the fine 
details of the surface brightness profile while kinematics have only a little 
impact. In this case, it is possible to obtain a best fit to the data in which 
the distribution of velocities is poorly reproduced.
However, differences in the projected density can be also produced 
by other factors (e.g. small variations in the the shape of the adopted 
distribution function, tidal effects, etc.) which are out of control.

To overcome to this problem, we constructed parametric models of rotating stellar 
systems by assuming {\it a priori} a density profile and deriving kinematics 
from the Jeans equation, assuming spherical symmetry.
Briefly, for an adopted density profile the associated profile of the second 
moment of the velocity $\langle v^{2} \rangle \equiv \langle v_{\phi}^{2} \rangle+
\langle v_{r}^{2}\rangle+\langle v_{\theta}^{2}\rangle$ has been calculated 
using the Jeans equation in spherical coordinates (eq. \ref{jeans_eq}; here $v_{r}$, $v_{\theta}$ and $v_{\phi}$ are the velocity component in the radial, polar and 
azimuthal directions\footnote{In this notation $v_{\phi}$ corresponds to the rotational velocity.}, respectively).
The relative contribution of ordered and random motions to $\langle v^{2}\rangle$ 
at a given position inside 
the cluster has been calculated adopting a parametric relation depending on 
the distance to the rotation axis. 
\begin{equation}
\centering
f=\frac{\langle v_{\phi}\rangle^{2}}{\langle v_{r}^{2}\rangle+\langle v_{\theta}^{2}\rangle+\langle v_{\phi}^{2}\rangle}=\frac{b}{3}~\frac{exp(R/R_{0})-1}{exp(R/a~R_{0})+1}$$
\end{equation}
where $b$ governs the strength of rotation, $R_{0}$ is a scale radius at 
which rotation approaches its maximum contribution to the kinetic energy and $a$ is a dampening factor at large radii.
A comprehensive description of the modelling technique is provided in Appendix B.

Of course, the use of the Jeans equation does not ensure self-consistency (i.e. 
it is possible to obtain a model with a corresponding distribution function 
which is negative in some point of the energy-angular momentum space). Moreover,
while spherical rotating models can exist \citep{lyndenbell60}, deviations from 
spherical symmetry have been observed in almost all GCs \citep{white87,chen10}.
The models adopted here thus provide only an empirical representation of the 
rotation of GCs which is supported by a general physical justification.

{The advantage of the above technique is that the density and velocity distribution 
profiles depend on non-degenerate sets of parameters. The fitting procedure has 
been performed in two subsequent steps: first,
for each cluster we adopted the 3D density profile of the best 
fit \citet{king66} model \citep[defined by the $W_{0}$ and $r_{c}$ parameters 
from][]{mclaughlin05} and we then searched the  
combination of parameters ($a,b,R_{0}$) which maximize the 
likelihood of eq. \ref{lik2_eq}, where the values of $\overline{v_{j,i}}$ and 
$\sigma_{j,i}$ were calculated from eq.s \ref{modproj_eq}.
A Markov-Chain Monte Carlo has been used to survey the parameter space searching 
for the parameters $a,~b$ and $R_{0}$ providing the maximum likelihood. 
The position ($\theta_{0}$) and inclination ($i$) angles or the rotation axis 
have been fixed to those calculated through the algorithm described in Sect. 
\ref{alg_sec}. Indeed, tests performed on the synthetic catalogs (see Sect. 
\ref{alg_sec}) indicate that the adopted algorithm provides unbiased estimates 
of these quantities while systematic offsets are obtained 
by leaving them as free parameters.

The best fit models for the 15 GCs of our final sample are shown in Fig. \ref{fit} and \ref{fit2}. 
In this Figures, the line-of-sight 
velocity dispersion and rotation profiles have been computed (for visualization purposes only) by binning the samples
of line-of-sight velocities and proper motions and applying the technique described in 
Sect. \ref{alg_sec} to individual bins. The distribution of velocities in the 
three components are plotted as a function of the distance along ($\Delta Y$) 
and from ($\Delta X$) the rotation axis

\begin{eqnarray*}
\Delta X&=&-R~sin(\theta-\theta_{0})\nonumber\\
\Delta Y&=&R~cos(\theta-\theta_{0})\nonumber\\
\nonumber\\
\end{eqnarray*}

 From Fig. \ref{fit} and \ref{fit2} it is apparent that while the considered models 
provide a good fit to the density and rotational/dispersion velocity profiles 
in most GCs of our sample, significant discrepancies are noticeable for a few 
clusters (e.g. NGC 6656 and Ter 5). 
Note however that in these GCs the mismatch between models and data 
is confined to the outer bins containing only a few stars. Instead, the model parameters 
are fitted using individual velocities so that the strongest constraint is 
provided by regions where more data points are available.

Usually, the ratio between the projected rotation amplitude and the central 
line-of-sight velocity dispersion ($v_{\phi,max}/\sigma_{0}$) is used as an indicator of the relative 
importance ofrotation \citep{davies83}. However, the use of this ratio has many 
drawbacks since it does not account for the different cluster concentrations and 
different shapes of the rotation curve \citep{binney05}.
To overcome this limitation and to assess the strength of rotation for each cluster, 
we calculated the fraction of kinetic energy in 
rotational motions $\xi\equiv\langle v_{\phi}^{2}\rangle/\langle \sigma^{2}\rangle$ 
of the corresponding best fit model. These fractions are listed in Table 2.
Note that $\xi$ is a global parameter and it is therefore a more robust 
rotation indicator in comparison with other 
estimators like e.g. $v_{\phi,max}/\sigma_{0}$, which are measured at different distances 
from the center, depend on the binning and use small subsamples of stars.

Moreover, we also calculated the model rotation period at the half-mass radius
as
\begin{equation}
P_{rot}=\frac{2\pi r_{h}}{v_{\phi}(r_{h})}
\label{per_eq}
\end{equation}

\section{Globular cluster parameter correlations}
\label{corr_sec}

\begin{figure*}
 \includegraphics[width=\textwidth]{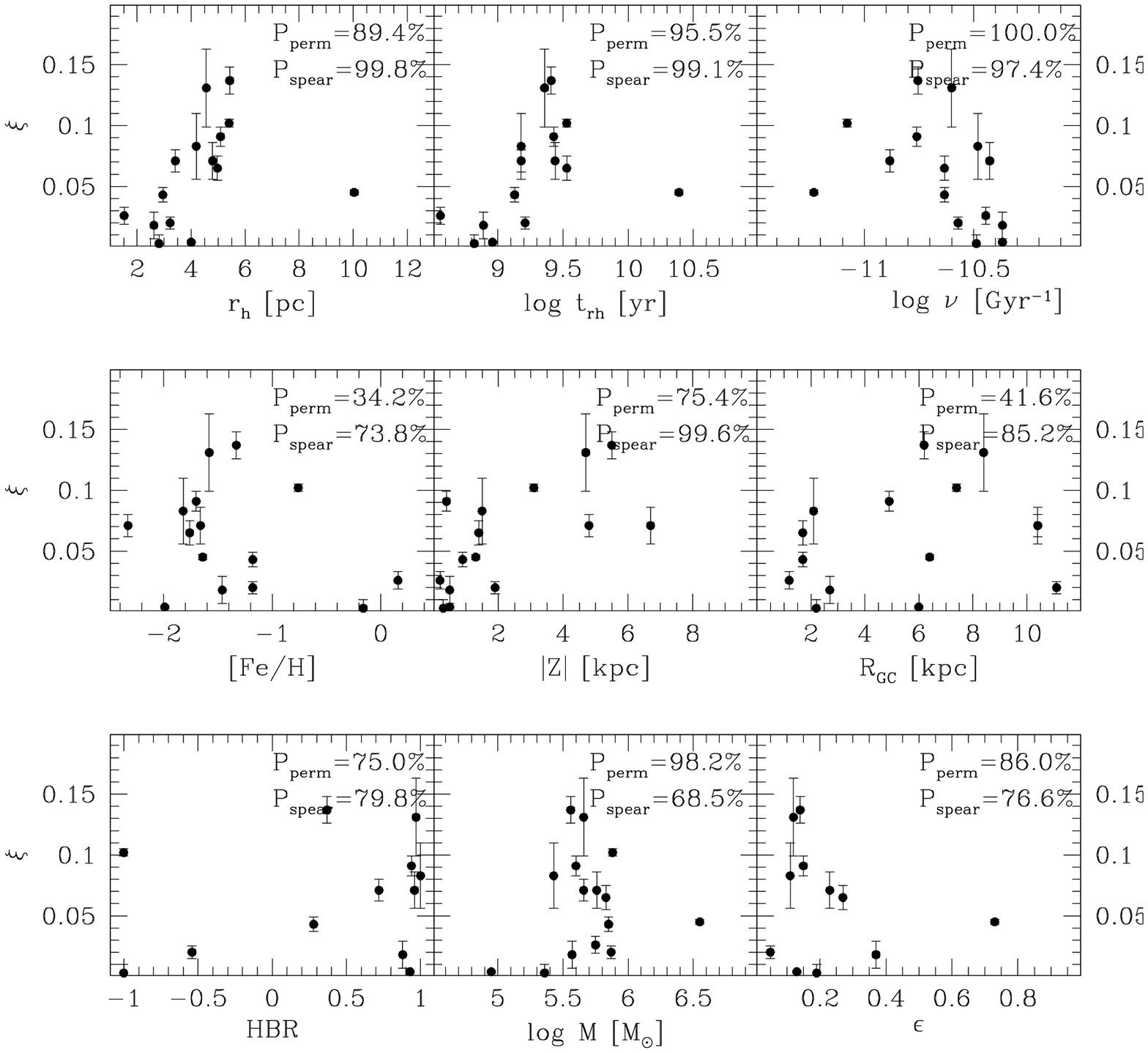}
 \caption{Correlation between the fraction of rotational kinetic energy ($\xi$) and various parameters. The
 correlation probability is indicated in each panel.}
\label{corr}
\end{figure*}

\begin{figure*}
 \includegraphics[width=\textwidth]{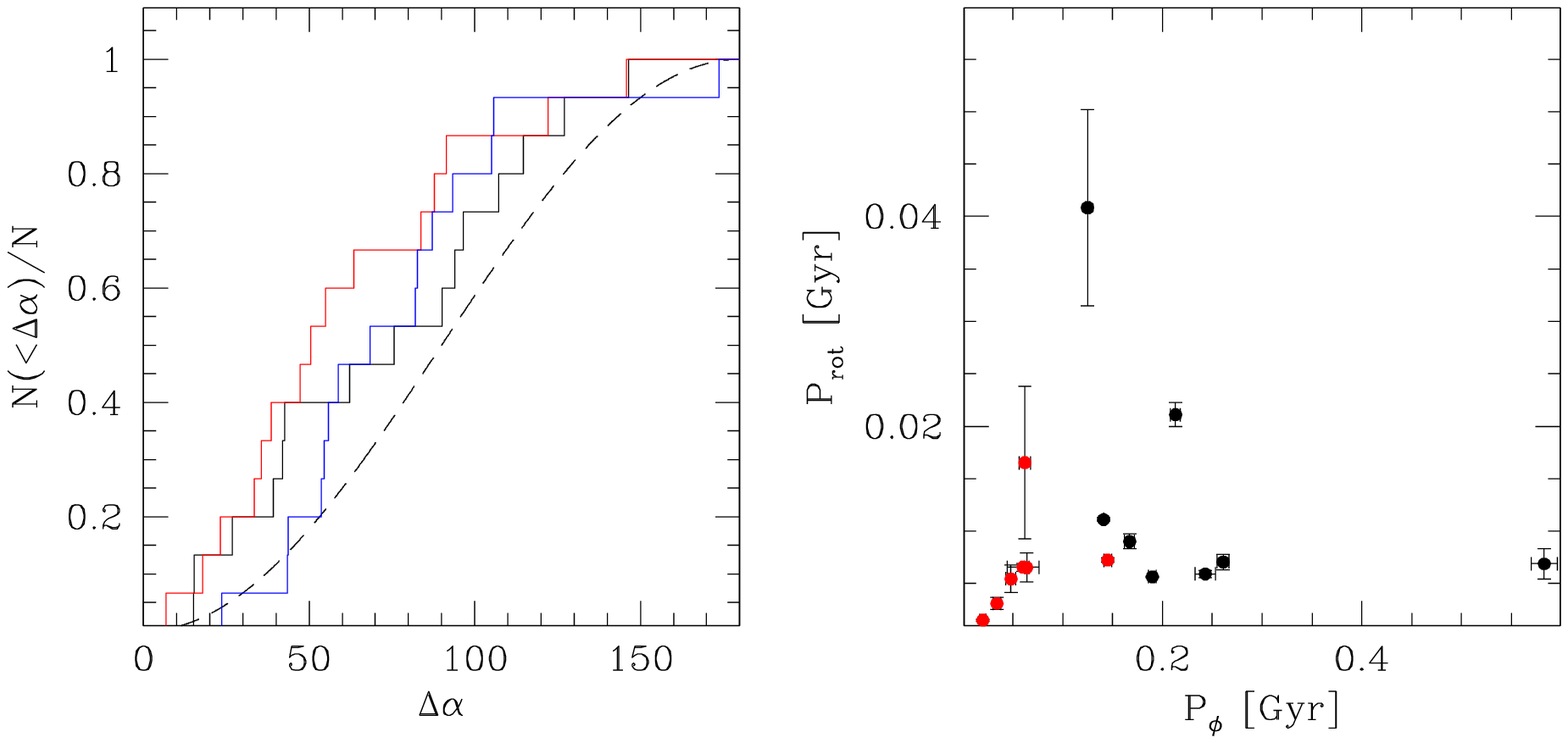}
 \caption{Left panel: cumulative distribution of the angle between the rotation axis and the 
 plane of the sky (black solid line), the Galactic pole (red line) and the present-day orbital 
 pole (blue line). The behaviour of a randomly oriented distribution is shown by the dashed line. 
 Right panel: comparison between the rotation period at the half-mass radius and 
 the orbital period. Red and black dots mark GCs with Galactocentric distances 
 smaller and larger than 6 kpc, respectively.}
\label{period}
\end{figure*}

The values of $\xi$ calculated in the previous Section have been 
used to search for correlations with other general and dynamical parameters.
We considered the horizontal branch morphology parameter \citep[$HBR$; from][]{lee94}, Galactocentric 
distance, height above the Galactic plane 
\citep[$R_{GC}$, $Z$; from][]{harris96} mass, half-mass radius, 
half-mass relaxation time \citep[$M$, $r_{h}$, $t_{rh}$; from][]{baumgardt18} 
metallicity \citep[Fe/H; from][]{carretta09}, ellipticity 
\citep[$\epsilon$; from][]{chen10} and destruction rate 
\citep[$\nu$ defined as the inverse of the time needed by the cluster to completely dissolve; from][]{baumgardt19}. 
Ellipticities have been de-projected using the technique described in \citet{cappellari07} and assuming the same 
inclination angles of the rotation axis, 
based on the hypothesis that rotation and flattening are linked.

To evaluate the significance of such 
correlations we performed both the Spearman rank correlation test and a
permutation test for Pearson's weighted correlation coefficient. For the latter test,
we assume $\xi$ as the dependent variable and
constructed $10^{4}$ random permutation of each independent variable.
For each set the weighted Pearson's correlation coefficient has been calculated.
The fraction of randomized sets with a correlation coefficient smaller than the 
one measured on the original data gives the probability that the two variables 
are correlated. 

The entire set of correlations is shown in 
Fig. \ref{corr} together with the corresponding correlation probabilities.
Unfortunately, given the small size of our database, we are not able to confirm 
or exclude any of the considered correlations at a significant confidence level.
On average, both the performed tests provide similar results with the Spearman 
test giving generally larger significance levels than the permutation one.
The largest correlation probabilities (still below the 3$\sigma$ level) are those 
with the half-mass radius and relaxation time, in agreement with what found by 
\citet{kamann18} and \citet{bianchini18}. Also noticeable is the anticorrelation with 
the destruction rate.
An interesting outlier in the above planes is NGC 5139 showing a relatively 
small rotation strength in spite of its long half-mass relaxation time and low 
destruction rate.

A relatively small correlation probability is 
instead that with the ellipticity. To further investigate the relation between 
rotation and flattening, we also calculated the angle in the plane of the sky 
between the position angle of the rotation axis with the direction of the
isophotal minor axis \citep[from][]{chen10}. A one-tailed Kolmogorov-Smirnov indicates a 
probability of 6.2\% that such a correlation occurs by chance.
It is worth noting that the ellipticity of \citet{chen10}, based on star counts 
in infrared images, are less affected by foreground extinction than the optical 
data, but more affected by individual star count numbers and by features
at larger radius such as tidal tails or outer stretching of the
cluster distribution. So, we also used the ellipticities provided by 
\citet{white87} which are based on the overall integrated light profiles and are 
therefore more representative of the (inner) brightest parts of the clusters. 
With this dataset we find a more significant correlation probability (99.8\% and 
85.7\% according to the permutation and the Spearman tests, respectively) although 
the alignment of the minor and rotation axes is not significant (11.6\% that 
the angle difference is extracted from a random distribution).

An interesting test for the impact of the tidal field on internal rotation 
has been done by calculating the angle between the rotation axes and the orbital 
poles of the GCs in our sample. For this purpose, we computed the orientation 
of the axes in the Galactic reference system and calculated the angles with 
the Galactic North (South) pole if the cluster has a positive (negative) angular 
momentum along the direction perpendicular to the Galactic plane 
\citep[from][]{gaia18b}.
Indeed, all the 15 GCs of our sample move in a region of the Galactic potential which 
is dominated by the disk, and their orbital poles oscillate
during the cluster orbit around the Galactic pole. Therefore, the average 
orientation of the orbital pole over many orbits is always pointed toward 
the Galactic pole. A one-tailed Kolmogorov-Smirnov indicates a 
probability of 1.5\% that such an angle is distributed following a constant 
probability per solid angle corresponding to a random orientation ($P(\Delta i)\propto 
sin~\Delta i$).
We also calculated the angle between the rotation axes and the instantaneous 
orbital poles, defined as the 
normal vectors to the planes containing the present-day position and velocity 
vectors. Also in this case, we found a probability of 11.3\% that such an angle 
comes from a random orientation of the rotation axes.
For comparison, if we perform the same test on the inclination angle with respect to the plane of the sky, 
and neglecting the dependence of the detection efficiency on it,
the probability that it is randomly distributed is 19.4\% (see the left panel of Fig. \ref{period}).

Finally, in the right panel of Fig. \ref{period} we compare the rotation period at the half-mass 
radius with the cluster orbital period \citep[from][]{gaia18b}.
While these two variables are not significantly correlated when considering 
the entire sample (the permutation and Spearman tests give probabilities of 
70.5\% and 92.0\%, respectively), a strong correlation of these two variables 
is apparent for GCs with $R_{GC}<6~kpc$. 
A least-square fit to this subsample of clusters gives a ratio 
$P_{\phi}/P_{rot}=9.2\pm2.5$.
Note however that, as it is immediately 
apparent from eq. \ref{per_eq}, the rotation period at the half-mass radius is 
proportional to the half-mass radius itself. On the other hand, it is well known 
that a half-mass radius vs. Galactocentric 
distance relation is present among Galactic GCs \citep{vandenbergh91}, with the GCs 
at large distances from the Galactic centre (i.e. those which take a long time to 
complete their orbits) being on average more extended.
So, since both the orbital and the rotational periods depend on the Galactocentric 
distance, the correlation between these two timescales
could be spurious. Indeed, this correlation is not associated 
with an alignment of the rotation axes with the orbital poles: if we restrict 
the sample to GCs with $R_{GC}<6$ kpc, the probability that the angle between these 
two directions is randomly distributed increases to 2.4\%.

\section{summary}
\label{concl_sec}

We constructed the most extensive set of kinematic information for stars in 62 Galactic
GCs, sampling $\sim40\%$ of the GC system of the Milky Way, matching Gaia proper
motions with the most comprehensive survey of line-of-sight velocities.
We explored for each analysed cluster the velocity distribution in the three components searching for
statistically significant signals of rotation.
We found robust evidence of rotation in 15 GCs of our sample at an amplitude
which cannot be explained by neither random nor systematic errors. For 9 more
GCs we found a signal of rotation mainly in the plane of the sky at a level
below the claimed amplitude of systematic uncertainties possibly present in the
Gaia catalog. Although the presence of a genuine rotation is well possible in these GCs, we 
cannot exclude that this evidence might be spuriously produced by the patchy 
distribution pattern of systematics. The present analysis adds an important piece of
information to the recent work by \citet{bianchini18} since, taking advantage of
the information on the velocity along the line-of-sight, {\it i)} it has a higher
efficiency in detecting GCs rotating with large inclination angles, and {\it
ii)} it allows to determine the inclination of the rotation axis with respect to
the line-of-sight.

The relative strength of ordered over random motions ($\xi$) has been also calculated by
means of the comparison with dynamical models. The derived values of 
$\xi$ appear to weakly correlate with the half-mass relaxation
time, with the GCs with longer relaxation times rotating faster. This
evidence, already noticed by \citet{kamann18} and \citet{bianchini18}, is in
agreement with the predictions of N-body simulations \citep{tiongco17}. 
It could suggest a primordial origin for the rotation of these 
stellar systems which is progressively erased by both internal and external 
dynamical processes. 
Indeed, two-body relaxation tend to randomize the orbits of stars thus erasing the
effect of ordered motions. In this picture, less evolved GCs (i.e. those with a
longer half-mass relaxation time) still maintain the evidence of their 
original rotation. Moreover, the observed anticorrelation between 
$\xi$ and the destruction rate suggests that GCs 
subject to a fast destruction process lost much of their original angular 
momentum. 
The massive GC NGC 5139 presents a remarkably 
small rotation strength in spite of its long half-mass relaxation time and low 
destruction rate. This finding suggests peculiar initial conditions for this 
stellar system characterised by a relatively small primordial rotation.

We also checked the importance of the Galactic tidal field in determining the
rotation of GCs. The comparison between the
orientation of the rotation axis with the average and the 
instantaneous orbital pole does not provide any significant conclusion.
It is interesting to notice the strong correlation between the rotation period
at the half-mass with the orbital period of those GCs at small Galactocentric
distances ($R_{GC}<6~kpc$). However, although a synchronization of internal and 
orbital rotation
is predicted by simulations \citep[][; who found however a much smaller ratio 
$P_{\phi}/P_{rot}\sim0.5$ 
close to the tidal radius]{tiongco16}, this evidence could be spuriously
produced by the half-mass radius vs. Galactocentric distance relation present in the Galactic 
GC system \citep{vandenbergh91}. This is also suggested by the lack of
any significant alignment of the rotation axis with the orbital poles in this
subsample of clusters. Note that tidal effects are expected to be effective in
creating a retrograde rotation only at large distances from the cluster centre
\citep{vesperini14}, while our data sample mainly the inner region of GCs. 
Therefore, the lack of a clear correlation between the rotation properties of
our GCs and the strength of the tidal field is not surprising. In this context,
the rotation found in the GCs of our sample is more likely reminiscent of their 
initial conditions.

At odds with what was found by \citet{fabricius14} and \citet{kamann18}, we do not 
find evidence of a link between rotation and flattening, in terms of 
neither the correlation between the rotation strength and the de-projected
ellipticity nor the alignment of the rotation axis with the isophotal minor axis. 
The lack of such a correlation, already reported by \citet{bellazzini12} and 
\citet{lardo15}, could be due to the effect of anisotropy and tidal distorsions 
in shaping the outermost regions of the clusters which introduce a spread in 
the $\xi$ diagram \citep[see ][]{kamann18}.

Unfortunately, all the conclusions drawn on the basis of the explored correlations suffer 
from the small size of our sample. In this situation, it is well possible 
that some of the explored correlations are actually real but remain hidden in 
the Poisson noise resulting statistically non-significant.
An important improvement is expected in the 
near future when the next Gaia releases will be available. According to the 
performance prediction of the Gaia consortium, the end of mission 
accuracy should improve by a factor of two and the amplitude of systematics 
is expected to significantly decrease. In that condition, the same analysis 
performed here should be able to clarify the presence of rotation in the 
GCs with uncertain detections and to 
construct a much larger sample of rotating GCs 
which can allow to verify the correlations analysed in this work.

\section*{Acknowledgments}
We warmly thank Michele Bellazzini and Paolo Bianchini for useful discussions. 
We also thank the anonymous referee for his/her helpful comments and 
suggestions.

\onecolumn
\appendix

\section{Derivation of the rotation velocity components}

Consider a reference frame defined such that a cluster rotates clockwise 
in the x-y plane with the z-axis directed in toward the direction of the
angular momentum.
The systemic velocities along the three components can be written as
\begin{eqnarray*}
v_{x}&=&\omega y\nonumber\\
v_{y}&=&-\omega x\nonumber\\
v_{z}&=&0\nonumber\\
\end{eqnarray*}

where $\omega\equiv\omega(x,y,z)$ is the angular velocity.
The velocity components measured by an observer looking at the cluster from an
inclined perspective ($v_{X}$, $v_{Y}$, $v_{Z}$) can be obtained by
sequentially applying two rotations along the x and z axes by angles $i$ and
$\theta_{0}$, respectively

\begin{align}
v_{X}&=&v_{x} cos\theta_{0}-v_{y} sin\theta_{0}cos i+v_{z}~sin\theta_{0}sin i&=&\omega~(x~sin\theta_{0} cos~i+y~cos\theta_{0})\nonumber\\
v_{Y}&=&v_{x} sin\theta_{0}+v_{y} cos\theta_{0}cos i-v_{z}~cos\theta_{0}sin i&=&-\omega~(x~cos\theta_{0} cos~i-y~sin\theta_{0})\nonumber\\
v_{Z}&=&v_{y} sin i+v_{z}~cos i                                              &=&-\omega~x~sin~i\nonumber\\
\label{transf1_eq}
\end{align}

Defining the position angle $\theta$ anti-clockwise from the Y axis we have
$$X=-R~sin\theta~~~~~~~Y=R~cos\theta$$
where $R=\sqrt{X^{2}+Y^{2}}$ is the projected distance from the cluster centre.
The coordinate transformation between the two reference systems are

\begin{align}
x&=&X cos\theta_{0}+Y sin\theta_{0}                     &=&-R~sin(\theta-\theta_{0})\nonumber\\
y&=&-X sin\theta_{0} cos~i+Y cos\theta_{0} cos~i+Z sin~i&=&R~cos(\theta-\theta_{0}) cos~i+Z sin~i\nonumber\\
z&=&X sin\theta_{0} sin~i -Y cos\theta_{0} sin~i+Z cos~i&=&-R cos(\theta-\theta_{0}) sin~i+Z cos~i\nonumber\\
\label{transf2_eq}
\end{align}

Consider the
projections of the velocity vector in the plane of the sky in the directions 
parallel and perpendicular to the rotation axis
\begin{eqnarray}
v_{\parallel}&=&-v_{X} sin\theta_{0}+v_{Y} cos\theta_{0}\nonumber\\
v_{\perp}&=&v_{X} cos\theta_{0}+v_{Y} sin\theta_{0}\nonumber\\
\label{transf3_eq}
\end{eqnarray}

Combining eq.s \ref{transf1_eq}, \ref{transf2_eq} and \ref{transf3_eq} we
finally find

\begin{eqnarray*}
v_{Z}&=&\omega~R~sin(\theta-\theta_{0}) sin~i\nonumber\\
v_{\parallel}&=&\omega~R~sin(\theta-\theta_{0}) cos~i\nonumber\\
v_{\perp}&=&\omega~[R~cos(\theta-\theta_{0}) cos~i+Z sin~i]\nonumber\\
\end{eqnarray*}

\section{parametric Jeans models of rotating spherical systems}

Consider the Jeans equation in spherical polar coordinates ($r,~\theta,~\phi$)

\begin{eqnarray}
\frac{\delta \rho\langle v_{r}^{2}\rangle}{\delta r}+\frac{1}{r}\frac{\delta\langle v_{r}v_{\theta}\rangle}{\delta \theta}+
\frac{\rho}{r}(2\langle v_{r}^{2}\rangle-\langle v_{\theta}^{2}\rangle-\langle v_{\phi}^{2}\rangle+\langle v_{r} v_{\theta}\rangle cot \theta)&=&-\rho\frac{\delta\Phi}{\delta r}\nonumber\\
\frac{\delta \langle v_{r}v_{\theta}\rangle}{\delta r}+\frac{1}{r}\frac{\delta \rho\langle v_{\theta}^{2}\rangle}{\delta\theta}+
\frac{\rho}{r}[3\langle v_{r}v_{\theta}\rangle+(\langle v_{\theta}^{2}\rangle-\langle v_{\phi}^{2}\rangle)cot\theta]&=&-\frac{\rho}{r}\frac{\delta\Phi}{\delta\theta}\nonumber\\
\label{jeans_eq}
\end{eqnarray}

where $\rho$ is the 3D density, $v_{r}$, $v_{\theta}$ and $v_{\phi}$ are the 
velocities along the three components, $r$ is the distance from the cluster 
centre and $\Phi$ is the gravitational potential.

An obvious solution can be found by assuming
\begin{eqnarray}
\langle v_{r} v_{\theta}\rangle&=&0\nonumber\\
\langle v_{r}^{2}\rangle=\langle v_{\theta}^{2}\rangle&=&\langle v_{\phi}^{2}\rangle\nonumber\\
\frac{\delta\rho}{\delta\theta}=\frac{\delta\Phi}{\delta\theta}=\frac{\delta \langle v_{r}^{2}\rangle}{\delta\theta}&=&0\nonumber\\
\label{iso_eq}
\end{eqnarray}

Thus, eq. \ref{jeans_eq} reduces to the spherical Jeans equation

\begin{equation}
\frac{\delta\rho\langle v_{r}^{2}\rangle}{\delta r}=-\rho\frac{\delta\Phi}{\delta r}
\label{jeans2_eq}
\end{equation}

The second velocity moments can be written separating the contribution of random
and systematic motions assuming

\begin{eqnarray}
\langle v_{r}^{2}\rangle&=&\sigma_{r}^{2}\nonumber\\
\langle v_{\phi}^{2}\rangle&=&\sigma_{\phi}^{2}+\langle v_{\phi}\rangle^{2}\nonumber\\
\label{mom_eq}
\end{eqnarray}

We adopted an empirical relation linking the fraction of kinetic 
energy in rotational motion as a function of the distance to the rotation axis. 

\begin{equation}
\centering
f=\frac{\langle v_{\phi}\rangle^{2}}{\langle v_{r}^{2}\rangle+\langle v_{\theta}^{2}\rangle+\langle v_{\phi}^{2}\rangle}=\frac{b}{3}~\frac{exp(R/R_{0})-1}{exp(R/a~R_{0})+1}$$
\label{f_eq}
\end{equation}

where $b$ governs the strength of rotation, $R_{0}$ is a scale radius at 
which rotation approaches its maximum contribution to the kinetic energy and $a$ is a dampening factor at large radii.

From eq.s \ref{iso_eq}, \ref{mom_eq} and \ref{f_eq} we have
\begin{eqnarray}
\sigma_{\phi}^{2}&=&(1-3f)~\sigma_{r}^{2}\nonumber\\
\langle v_{\phi}\rangle^{2}&=&3f~\sigma_{r}^{2}\nonumber\\
\label{rot_eq}
\end{eqnarray}

For a given density profile, the potential derivative is given by the gravitational 
acceleration

$$ \frac{\delta\Phi}{\delta r}=-\frac{4\pi G}{r^{2}} \int_{0}^{r}\rho(r') r'^{2} dr'$$

while the velocity dispersions in the $r$ and $\phi$ components, as well as the 
mean rotational velocity can be calculated from eq.s \ref{jeans2_eq} and 
\ref{rot_eq}.

For a given density profile and a combination of parameters $(a,~b,~R_{0})$ the ratio between 
rotational and overall kinetic energy is given by
$$\xi\equiv\frac{\langle \langle v_{\phi}\rangle^{2} \rangle}{\langle 3~\sigma_{r}^{2}\rangle}=\frac{\int_{0}^{r_{t}} \int_{0}^{\sqrt{r_{t}^{2}-z^{2}}} R \rho f \sigma_{r}^{2} dR dz}{\int_{0}^{r_{t}} \int_{0}^{\sqrt{r_{t}^{2}-z^{2}}} R \rho \sigma_{r}^{2} dR dz}$$

The mean velocities and dispersions along the projected components ($\perp,~\parallel,~LOS$) can be 
calculated as a function of the inclination angle $i$ 

\begin{eqnarray}
\langle v_{\perp} \rangle&=&\frac{1}{\Sigma}\int_{-\infty}^{+\infty} \rho~\langle v_{\phi}\rangle\frac{y~cos~i+z~sin~i}{\sqrt{(y~cos~i+z~sin~i)^{2}+x^{2}}}dz\nonumber\\
\langle v_{\parallel} \rangle&=&-\frac{1}{\Sigma}\int_{-\infty}^{+\infty} \rho~\langle v_{\phi}\rangle\frac{x~cos~i}{\sqrt{(y~cos~i+z~sin~i)^{2}+x^{2}}}dz\nonumber\\
\langle v_{LOS} \rangle&=&-\frac{1}{\Sigma}\int_{-\infty}^{+\infty} \rho~\langle v_{\phi}\rangle\frac{x~sin~i}{\sqrt{(y~cos~i+z~sin~i)^{2}+x^{2}}}dz\nonumber\\
\sigma_{\perp}^{2}&=&\frac{1}{\Sigma}\int_{-\infty}^{+\infty} \rho~\sigma_{r}^{2}dz-\langle v_{\perp}\rangle^{2}\nonumber\\
\sigma_{\parallel}^{2}&=&\frac{1}{\Sigma}\int_{-\infty}^{+\infty} \rho~\sigma_{r}^{2}dz-\langle v_{\parallel}\rangle^{2}\nonumber\\
\sigma_{LOS}^{2}&=&\frac{1}{\Sigma}\int_{-\infty}^{+\infty} \rho~\sigma_{r}^{2}dz-\langle v_{LOS}\rangle^{2}\nonumber\\
\label{modproj_eq}
\end{eqnarray}

where 

$$\Sigma=\int_{-\infty}^{+\infty} \rho dz$$

is the projected density, x is directed orthogonal to the projection of the rotation axis into the plane of the sky, z 
is the distance along the line of sight, and
$\langle v_{\phi}\rangle$, $\sigma_{r}^{2}$ are 
calculated at intrinsic coordinates x',y' and z'

\begin{eqnarray*}
x'&=&x\nonumber\\
y'&=&y~cos i+z~sin i \nonumber\\
z'&=&-y~sin i+z~cos i\nonumber\\
\end{eqnarray*}

where z' is the height above the equatorial plane and $R=\sqrt{x'^{2}+y'^{2}}$ is the distance from the rotation axis.

\label{lastpage}

\end{document}